\documentclass[mreferee]{gjimod}
\usepackage{footnote}
\usepackage{algorithm}
\usepackage{algorithmic}
\usepackage[pdftex]{graphicx}
\usepackage{epstopdf}
\usepackage{epsfig,subfigure}
\usepackage{epsf}
\usepackage{graphics}
\usepackage{url}
\usepackage{xcolor}
\usepackage{amssymb}
\usepackage{amsmath}

\newcommand{\SNR}{\mathrm{SNR}}

\newcommand{\bfd}{\mathbf{d}}
\newcommand{\bfdo}{\mathbf{d}^{\mathrm{obs}}}
\newcommand{\bfde}{\mathbf{d}^{\mathrm{exact}}}
\newcommand{\bfm}{\mathbf{m}}
\newcommand{\bfma}{\bfm^{\mathrm{apr}}}
\newcommand{\Wdepth}{W_{\mathrm{depth}}}
\newcommand{\Wd}{W_{\bfd}}
\newcommand{\Wh}{W_{\mathrm{h}}}
\newcommand{\WL}{W_{{\mathrm{L_p}}}}
\newcommand{\Gg}{G_{\textrm{1}}}
\newcommand{\Gm}{G_{\textrm{2}}}
\newcommand{\Gi}{G_{\textrm{i}}}
\newcommand{\qg}{q_{\textrm{1}}}
\newcommand{\qm}{q_{\textrm{2}}}
\newcommand{\alphag}{\alpha_{\textrm{1}}}
\newcommand{\alpham}{\alpha_{\textrm{2}}}
\newcommand{\betag}{\beta_{\textrm{1}}}
\newcommand{\betam}{\beta_{\textrm{2}}}
\newcommand{\betai}{\beta_{\textrm{i}}}
\newcommand{\chig}{\chi_{\textrm{1}}}
\newcommand{\chim}{\chi_{\textrm{2}}}
\newcommand{\chii}{\chi_{\textrm{i}}}
\newcommand{\mg}{m_{\textrm{1}}}
\newcommand{\mm}{m_{\textrm{2}}}

\newcommand{\bfdg}{\bfdo_\textrm{1}}
\newcommand{\bfdm}{\bfdo_\textrm{2}}
\newcommand{\bfdi}{\bfdo_\textrm{i}}

\newcommand{\bfdei}{\bfde_\textrm{i}}
\newcommand{\bfmg}{\bfm_\textrm{1}}
\newcommand{\bfmm}{\bfm_\textrm{2}}
\newcommand{\bfmi}{\bfm_\textrm{i}}
\newcommand{\bfmag}{\bfma_\textrm{1}}
\newcommand{\bfmam}{\bfma_\textrm{2}}
\newcommand{\bfmai}{\bfma_\textrm{i}}
\newcommand{\Dg}{D_{\textrm{1}}}
\newcommand{\Dm}{D_{\textrm{2}}}
\newcommand{\Wg}{W_{\textrm{1}}}
\newcommand{\Wm}{W_{\textrm{2}}}
\newcommand{\Wi}{W_{\textrm{i}}}
\newcommand{\Wdepthg}{(\Wdepth)_{\textrm{1}}}
\newcommand{\Wdepthm}{(\Wdepth)_{\textrm{2}}}
\newcommand{\Wdepthi}{(\Wdepth)_{\textrm{i}}}
\newcommand{\Whg}{(\Wh)_{\textrm{1}}}
\newcommand{\Whm}{(\Wh)_{\textrm{2}}}
\newcommand{\Whi}{(\Wh)_{\textrm{i}}}
\newcommand{\WLg}{(\WL)_{\textrm{1}}}
\newcommand{\WLm}{(\WL)_{\textrm{2}}}
\newcommand{\WLi}{(\WL)_{\textrm{i}}}
\newcommand{\Wdg}{W_{{{\bfd}}_{\textrm{1}}}}
\newcommand{\Wdm}{W_{{{\bfd}}_{\textrm{2}}}}
\newcommand{\Wdi}{W_{{{\bfd}}_{\textrm{i}}}}

\newcommand{\SNRi}{\SNR_\mathrm{i}}

\newcommand{\bft}{\mathbf{t}}
\newcommand{\tx}{\bft^x}
\newcommand{\ty}{\bft^y}
\newcommand{\tz}{\bft^z}

\newcommand{\bfx}{\mathbf{x}}

\newcommand{\bfB}{\mathbf{B}}
\newcommand{\bff}{\mathbf{f}}

\newcommand{\Kmax}{\textrm{MAXIT}}

\newcommand{\Rmn}[2]{\mathcal{R}^{#1\times #2}}
\newcommand{\Rm}[1]{\mathcal{R}^{#1}}
\newcommand{\sparse}{\texttt{sparse }}
\newcommand{\diag}{\mathrm{diag}}
\newcommand{\blkdiag}{\mathrm{block\_diag}}
\newcommand{\blkstack}{\mathrm{block\_stack}}

\title[Generalized L$_p$-norm joint inversion]{Generalized L$_p$-norm joint inversion of gravity and magnetic data using cross-gradient constraint}
\author[S. Vatankhah, S. Liu, R.~A. Renaut, X. Hu, M. Gharloghi]{Saeed Vatankhah $^1$$^,$$^2$ ,  Shuang Liu $^1$,  Rosemary A. Renaut $^3$, Xiangyun Hu $^1$, \and  Mostafa Gharloghi $^2$ \\
$^1$ Hubei Subsurface Multi-scale Imaging Key Laboratory, Institute of Geophysics and Geomatics,\\
 China University of Geosciences, Wuhan, China\\
$^2$ Institute of Geophysics, University of Tehran, Tehran, Iran\\
$^3$ School of Mathematical and Statistical Sciences, Arizona State University, Tempe, AZ, USA\\
}
\begin{document}
\maketitle

\begin{summary}
A generalized unifying approach for $L_{p}$-norm joint inversion of gravity and magnetic data using the cross-gradient constraint is presented. The presented framework incorporates stabilizers that use $L_{0}$, $L_{1}$, and $L_{2}$-norms of the model parameters, and/or the gradient of the model parameters.  Furthermore, the formulation is developed from standard approaches for independent inversion of single data sets, and, thus, also facilitates the inclusion of necessary model and data weighting matrices that provide, for example, depth weighting and imposition of hard constraint data. The developed efficient algorithm can, therefore, be employed to provide  physically-relevant smooth, sparse, or blocky target(s) which are relevant to the geophysical community.
 Here, the nonlinear objective function, that describes the inclusion of all stabilizing terms and the fit to data measurements, is minimized iteratively by imposing stationarity on the linear equation that results from applying   linearization of the objective function about a starting model. To numerically solve the resulting linear system, at each iteration, the conjugate gradient (CG) algorithm is used. The general framework is then validated for three-dimensional synthetic models for both sparse and smooth reconstructions, and the results are compared with those of individual gravity and magnetic inversions. It is  demonstrated that the presented joint inversion algorithm is practical and   significantly improves  reconstructed models obtained by independent inversion.
\end{summary}
\section{Introduction}\label{intro}
Potential field surveys, gravity and magnetic, have been reported as effective strategies for delineating  subsurface geological targets. They are  applied in a wide range of studies including, for example, oil and gas exploration, mining applications, and mapping the basement topography \cite{nabighian:2005,Blakely}. These surveys are relatively cheap, non-destructive passive remote sensing methods, and can provide valuable information of the subsurface targets.  Yet, they only require the measurement of variations in the Earth's natural fields that are caused by changes in the physical properties of the subsurface rocks.  In the interpretation process,  the acquired survey data can be used in an automatic inversion algorithm for estimation of specific parameters of a subsurface target, for example its geometry or physical properties. It is well known, however,  that the potential field inversion problem is ill posed.  Identifiability of  stable and physically-relevant solutions is then   obtained by application of suitable regularization strategies. An independent solution of the inverse problem for either gravity, or magnetic, data for the survey area will only provide information about the density or susceptibility, respectively, of the subsurface. On the other hand, complementary solution of the inverse problem for both data sets can be used to reveal  both density and magnetization variations present in a subsurface target. Thus, it is more appropriate to perform a simultaneous joint inversion that uses  both data sets.  Combined with regularization, this is an effective strategy for yielding a reliable subsurface geological model, that simplifies the interpretation of the subsurface target(s). Thus, the development of efficient and stable joint inversion algorithms has received increased attention in the geophysical community.

Many different techniques have been developed for the simultaneous joint inversion of geophysical data sets. Generally, these techniques can be categorized into two main groups: (i) petrophysical, and (ii) structural approaches. Petrophysical techniques rely on a direct relationship between two or more  physical properties of the subsurface target, for example the assumption that the resistivity and the velocity are both functions of porosity and water saturation \cite{GM:03}. Although, this strategy is attractive, it does depend on finding a reliable empirical relationship between physical properties. This is a difficult task for general geological media because there is usually no simple or single relationship that approximates the whole range of effects \cite{GM:03}. Further discussion of the details of the application of petrophysical techniques for joint inversion is provided in the literature, including for example in  Nielson \& Jacobsen \shortcite{NJ:00} and Moorkamp et al. \shortcite{Moorkamp:2011,Moorkamp:2013}. Structural approaches use, instead, the model topology in order to enhance the structural similarity of  reconstructed models \cite{HO:97,GM:03,GM:04,TL:06,FG:09}.  The main idea is that changes, at any point in the different models, should occur in the same or opposite spatial directions, or alternatively, changes will only occur in one of the models. Mathematically, this may be  achieved by forcing the cross product of  the gradient of the different model parameters to be  zero everywhere \cite{GM:03,GM:04,TL:06}. Indeed, many successful results for simultaneous joint inversion with the inclusion of the cross-gradient constraint  have been reported,  \cite{GM:03,GM:04,TL:06,Gallardo:07,FG:09,HG:13,FGG:15,Gross:19,ZhWa:19}. On the other hand,  Zhdanov et al. \shortcite{ZGW:12} observed that the Gramian constraint can be used to enhance the correlation between different physical properties and/or their attributes. In this approach,  the correlation is enhanced by minimizing the determinant of the Gram matrices of multi-model model parameters during the inversion process. The Gramian constraint approach is general; extant methods based on petrophysical correlations or cross-gradient minimization are special case reductions \cite{ZGW:12}. Moreover, unlike the case when performing petrophysical joint inversion, a priori information about the specific relationships between the different  physical parameters, or their attributes, is not required. The methodology has been used extensively in the joint inversion of different data sets, see for example Zhdanov \shortcite{Zhdanov:15}, Lin \& Zhdanov \shortcite{LZ:18} and Jorgensen \& Zhdanov \shortcite{JZ:19}. In this study we assume that the structure of the subsurface target(s) yields density and susceptibility model parameters over an approximately similar structure.  Thus, our focus is on the use of the cross-gradient constraint within a general L$_p$ formulation for efficient simultaneous joint inversion of  gravity and magnetic data sets.

Different types of stabilizers  have been adopted for the inversion of  potential field data, dependent on the desired smoothness, or otherwise, of model features that are to be recovered. For example, it may be appropriate to reconstruct a model which only represents the large-scale features of the subsurface under the survey area without any arbitrary discontinuities. This is achieved with the maximum smoothness stabilizer which uses a  L$_2$-norm\footnote[1]{The L$_p$-norm of a vector $\bfx \in \Rm{n}$ is defined as $ \| \bfx \|_{p}=(\sum _{i} ^{n} | x_{i}|^{p})^{1/p} $, $p \geq 1$, and $\| \bfx \|_{0}$ counts  the number of nonzero entries in $\bfx$.}  of  the  gradient  of  the  model  parameters, \cite{CPC:87,LiOl:96,Pi:97,LiOl:98}. On the contrary, when it is anticipated that the subsurface structure exhibits discontinuities, stabilization can be achieved by imposing the L$_1$-norm, or L$_0$-norm, on  the gradient  of  the  model  parameters \cite{FaOl:98,PoZh:99,BCO:2002,VRA:2018a,FoOl:19}.  Alternatively, when it can be assumed that the subsurface targets are localized and compact, it is more appropriate to apply the L$_1$ or L$_0$-norms directly on the model parameters \cite{LaKu:83,PoZh:99,BaSi:94,ZhTo:04,Ajo:2007,VAR:2014a,VRA:2014b,SunLi:2014,Zhdanov:15,VRA:2017}. In the potential field literature, stabilization by application of the L$_0$-norm on the model parameters  is usually referred  to as the compactness constraint, whereas application of the L$_1$-norm, or L$_0$-norm, on  the gradient is referred to as  total variation (TV) and minimum gradient support  (MGS), stabilization, respectively. A unifying approach for application of these constraints for single potential field inversion is presented in Vatankhah et al. \shortcite{VRL:19a}. This approach also includes the modification of the stabilizers to account for additional model and data weighting matrices, such as required for imposition of depth weighting and hard constraint conditions, \cite{LiOl:96,BoCh:2001}. Note that the use of depth weighting counteracts the natural rapid decay of the kernels, dependent on the specific kernel, whether magnetic or gravity.  This then facilitates the contribution of all prisms at depth to the surface measurements with an approximately equal probability through the inversion algorithm.  The hard constraint weighting is used to impose available geological information on the reconstructed model. Consequently, any practical algorithm for the joint inversion of gravity and magnetic data should also incorporate such weighting schemes. Here, we extend this unifying framework for simultaneous joint inversion of gravity and magnetic data sets in conjunction with the use of the cross-gradient constraint.  
 
The well-known and widely-used formula of Fregoso \& Gallardo \shortcite{FG:09} for the joint inversion of gravity and magnetic data, incorporating the  cross-gradient constraint, is based on the use computationally of the generalized nonlinear least-squares framework developed by Tarantola \& Valette \shortcite{TV:82}. Here, we wish to include deterministic constraints within the simultaneous joint inversion of the data and thus adopt a deterministic viewpoint for the parameter estimation. The nonlinear objective function, that describes the inclusion of all stabilizing terms and the fit to data measurements, is minimized iteratively by imposing stationarity on the linear equation that results from applying   linearization of the objective function about a starting model.  To perform the inversion, the iteratively re-weighted least-squares (IRLS) strategy  is then used \cite{WoRo:07}. At each iteration, the conjugate gradient (CG) algorithm is applied  to numerically solve the resulting linear system.

The paper organized as follow. In Section~\ref{JointInversionMethodology}, the theoretical development of the algorithm is presented, along with   a unifying framework that makes it  possible to combine  different types stabilizers within the context of  joint inversion. In Section~\ref{Synthetic}, the developed algorithm is validated on synthetic examples. Here, two synthetic models are used; and both sparse and smooth reconstructions are considered. Section~\ref{conclusion}  is dedicated to a discussion of conclusions and future topics for research.

\section{Joint inversion methodology}\label{JointInversionMethodology}
To formulate the problem, we use the well-known strategy for linear inversion of potential field data in which the subsurface is divided into a set of rectangular prisms with fixed size but unknown physical properties \cite{LiOl:96,BoCh:2001}. Here, it is assumed that there is no remanent magnetization, and that self-demagnetization effects are also negligible. For ease of exposition, we first introduce some basic notation for stacking of vectors (matrices) and generation of block diagonal matrices.   We use  $\blkstack \left(\cdot, \cdot \right)$ to indicate the stacking of vectors (or matrices) with the same number of columns in one vector (or matrix). Further,  $\blkdiag \left( A, B\right) $  indicates a block diagonal matrix of size $(m_A+m_B) \times (n_A+n_B)$ when $A$ and $B$ are of sizes $(m_A \times n_A)$ and $(m_B\times n_B)$, respectively. Both definitions extend immediately for more than two entities. 

We suppose that $m$ measurements are taken for two sets of potential field data \footnote[2]{We could assume different numbers of measurements for each field, $\mg$ and $\mm$ but for simplicity of the discussion we immediately assume $\mg=\mm=m$.}. These are the  vertical components of the gravity and total magnetic fields, and they are stacked in vectors $\bfdg$, and $\bfdm$, each of length $m$, respectively. The unknown physical parameters, the density and the susceptibility, of $n$ prisms are also stacked in vectors $\bfmg$, and $\bfmm$, respectively.
The data vectors and model parameters are then stacked consistently in vectors $\bfdo=\blkstack \left( \bfdg, \bfdm\right)  \in \Rm{2m}$, and  $\bfm=\blkstack \left( \bfmg, \bfmm \right) \in \Rm{2n}$.   The measurements are connected to the model parameters via  $G\bfm=\bfdo$ where $G= \blkdiag \left(\Gg,\Gm\right)  \in \Rmn{2m}{2n}$, and $\Gg$ and $\Gm$ are the linear forward modeling operators for gravity and magnetic kernels, respectively.  There are different alternative formulas which can be used to compute the entries of matrices $\Gg$ and $\Gm$. Here, we use the formulas developed by Ha{\'a}z \shortcite{Haaz}, for computing the vertical gravitational component, and  Rao \& Babu \shortcite{RaBa:91}, for the total magnetic field anomaly, of a right rectangular prism, respectively. 

The goal of the inversion is to find geologically plausible models $\bfmg$ and $\bfmm$ that predict $\bfdg$ and $\bfdm$, respectively, via a simultaneous joint algorithm that also facilitates the incorporation of relevant weighting matrices in the algorithm. We formulate the joint inversion for the determination of the model parameters  $\bfmg$ and $\bfmm$  as the minimization of the global objective function, in which parameters $\alpha$ and $\lambda$ are relative weighting parameters for the respective terms, 
\begin{align}\label{OriginalP}
P^{(\alpha,\lambda)}(\bfm)=\| \Wd (\bfdo-G \bfm)\|_2^2 + \alpha^{2}~\| WD(\bfm-\bfma) \|_2^2 +  \lambda^{2}~\| \bft(\bfm) \|_2^2. 
\end{align}
The data misfit term, $\| \Wd (\bfdo-G \bfm)\|_2^2$, measures how  well the calculated data reproduce the observed data. Diagonal matrix  $\Wd=\blkdiag\left( \Wdg, \Wdm \right) \in \Rmn{2m}{2m}$, where $\Wdg$ and $\Wdm$ are diagonal weighting matrices for the gravity and magnetic data, respectively. Here we suppose that these diagonal elements are the inverses of the standard deviations of the independent, but potentially colored, noise in the data. Zhang \& Wang \shortcite{ZhWa:19} considered an alternative weighting based on the individual row norms which  could also be used here. The stabilizer, $\| WD(\bfm-\bfma) \|_2^2 $, controls the growth of the solution with respect to the weighted norm, and is especially significant as it determines the structural qualities of the desired solution. Here, this  stabilizer is presented  through a general L$_{2}$-norm formulation, but  we will discuss how different choices of $W$ and $D$ lead to different L$_{p}$-norm stabilizations. In \eqref{OriginalP} the vector $\bfma = \blkstack\left( \bfmag, \bfmam \right) \in \Rm{2n}$ is an initial starting model that maybe known from previous investigations. It is also possible to set $\bfma =\mathbf{0}$.  The link between the gravity and magnetic models in this inversion algorithm is the cross-gradient function $ \bft(\bfm) \in \Rm{3n}$. For this study, we assume that the model structures for  $\bfmg$ and $\bfmm$ are approximately the same, and thus that it is important to measure the structural similarity using the cross-gradient constraint which will be approximately zero for   models with similar structures. We therefore use
\begin{align}\label{crossgradientfunction}
\bft(\bfm)=\nabla \bfmg~(x,y,z) \times  \nabla \bfmm~(x,y,z),
\end{align}  
where $\nabla$ indicates the gradient operator \cite{GM:03} and  structural similarity is achieved when $\bft(\bfm)= \mathbf{0}$, see Appendix~\ref{App1} for details. As noted already in Section~\ref{intro}, this corresponds to the case in which the gradient vectors are in the same or opposite direction, or, alternatively,  one of them is zero \cite{GM:03,GM:04,TL:06,Gallardo:07,FG:09,HG:13,FGG:15}. From a geological viewpoint this means that if a boundary exists then
it must be sensed by both methods in a common orientation
regardless of how the amplitude of the physical property
changes \cite{GM:03}. This means that information that is contained in one model is relevant to the other model and vice versa. Therefore,  structures determined by one model can assist with the identification of structures in the other model, and, as a consequence,  the structures of the two models can correct each other throughout the joint inversion process \cite{GM:03,HG:13}. On the other hand, while it is assumed that both models have similar structures at similar locations, it is also possible for one model to have a structure in a location where the other model has none \cite{HG:13}.  Further details about characteristic properties of the cross-gradient constraint are provided in   Gallardo \& Meju \shortcite{GM:03}, Tryggvason \& Linde \shortcite{TL:06}, and Fregoso \& Gallardo \shortcite{FG:09}.

The stabilizing  term, $\| WD(\bfm-\bfma) \|_2^2$, in \eqref{OriginalP} has a very significant impact on the solution that is obtained by minimizing \eqref{OriginalP}.  Depending on the type of desired model to be recovered, there are many choices that can be considered for this stabilization, and that have been extensively adopted by the geophysical community. Here,  we show how it is possible to use the given weighted L$_2$-norm regularizer to approximate different L$_{p}$-norm stabilizers, $ 0\leq p \leq 2$, \cite{VRL:19a}. Suppose $D$  is the identity matrix, $D=I_{2n}$, and $W$ is selected as  $ W=\blkdiag \left( \Wg,  \Wm \right) \in \Rmn{2n}{2n}$, in which,  
\begin{align}\label{Wprod}
\Wi={\Wdepthi} {\Whi} {\WLi} \in \Rmn{n}{n}, \quad \textrm{i} = 1, 2.
\end{align}
 Here, the diagonal weighting matrix ${\WLi} \in \Rmn{n}{n}$ is defined, assuming entries are calculated elementwise, by 
\begin{align}\label{WL1}
\WLi = \diag \left( 1 /((\bfmi-\bfmai)^{2}+\epsilon^{2})^{(2-p)/4}\right), \textrm{i} = 1, 2.
\end{align}
When  $p=0$ and $p=1$,  compact L$_{0}$-norm and L$_{1}$-norm solutions are obtained, respectively. The choice $p=2$ provides the L$_{2}$-norm solution of the model parameters. The parameter $\epsilon$ is a small positive number, $0< \epsilon \ll 1$, which is added to avoid the possibility of division by zero, and has an important effect on the solution. When $\epsilon$ is very small the solutions are sparse, while for large values the solutions are smooth. Further discussion on the impact of  $\epsilon$ is given,  for example, in Last \& Kubik \shortcite{LaKu:83}, Farquharson \& Oldenburg \shortcite{FaOl:98},  Zhdanov  \& Tolstaya \shortcite{ZhTo:04}, Ajo-Franklin et al. \shortcite{Ajo:2007}, Fiandaca et al. \shortcite{FDVA:15}, and Fournier \& Oldenburg \shortcite{FoOl:19}. On the other hand, it is also possible to chose $D$ to provide an approximation to the gradient of the model parameters. Suppose, for example, that $\Dg=\blkstack\left(D_x, D_y,D_z\right) \in \Rmn{3n}{n}$, where $D_{x}$, $D_{y}$, and $D_{z}$ are square and provide discrete approximations for derivative operators in $x$, $y$, and $z$-directions. Then, defining $\mathbf{0}_{3n \times n}$ to be the zero matrix of size $3n \times n$, and setting $\Dm=\Dg$, we can use the matrix 
\begin{align}\label{matrixD}
D=\left(  \begin{array}{c c}  \Dg &\mathbf{0}_{3n \times n}  \\ \mathbf{0}_{3n \times n} & \Dm
 \end{array} \right) \in \Rmn{6n}{2n},
\end{align}
so that $D\bfm$ yields the approximate gradient of $\bfm$. More details  about the structure of the matrices $D_{x}$, $D_{y}$, and $D_{z}$  can be found in  Li \& Oldenburg \shortcite{LiOl:2000}, Leli{\'e}vre \& Oldenburg \shortcite{LeOl:09}, and Leli{\'e}vre \& Oldenburg \shortcite{LeFa:13}. 
Then, with this definition for $D$, and again using  element-wise calculations, $\WLi$ in \eqref{WL1} is replaced by 
\begin{align}\label{WL2}
{\WLi} = \diag \left( 1 / \left( (D_{x}(\bfm-\bfmai))^{2}+(D_{y}(\bfm-\bfmai))^{2}+ (D_{z}(\bfm-\bfmai))^{2} + \epsilon^{2}\right)^{(2-p)/4} \right).
\end{align}
But now, for the multiplication in equation \eqref{OriginalP} to be dimensionally consistent, $W$ is replaced by
\begin{align}\label{matrixW}
W=\blkdiag\left(\Wg, \Wg,\Wg,\Wm,\Wm,\Wm\right) \in \Rmn{6n}{6n}.
\end{align}
In the definition of  $\WLi$, given in \eqref{WL2},  picking $p=2$ produces a solution with minimum structure yielding a smooth model without arbitrary discontinuities \cite{CPC:87,LiOl:96,Pi:97}. If we anticipate that there are true discontinuous jumps, then, it is possible to take  $p=1$ or $p=0$ for a  total variation (TV) or minimum gradient support (MGS) stabilization, respectively. In summary, all aforementioned definitions indicate how, dependent on the choices of $p$ and $D$ in $W$, it is possible to use the objective function \eqref{OriginalP} with a desired stabilizer. Well-known stabilizers, including TV, MGS, minimum structure, compactness, and L$_{1}$-norm  can all be incorporated in a joint inversion methodology. Moreover, this unifying framework allows the use of additional stabilizers, which are not common in potential field inversion, simply by changing the choice of $p$. 

In \eqref{Wprod}, the diagonal depth weighting matrix, $\Wdepthi= \diag(1/(z_{j}+z_{0})^{\betai})$ is used to counteract the rapid decay of the kernel with depth \cite{LiOl:96,Pi:97}. Here $z_{j}$ is the mean depth of prism $j$, $z_{0}$ depends both upon prism size and the  height of the observed data. With application of appropriate depth weighting, as determined by  parameter $\betai$, all prisms participate with an approximately equal probability in the inversion process. The diagonal hard constraint matrix $\Whi$, is generally an identity matrix. If geological information, or prior investigations in the survey area, can provide the values of the model parameter for some prisms, then, the information is included in $\bfmai$, and the corresponding diagonal entries of $\Whi$ are set to a large value \cite{BoCh:2001,VAR:2014a,VRA:2018b}. These known parameters are kept fixed during the iterative minimization of \eqref{OriginalP}. Equivalently,    the inversion algorithm searches only for unknown model parameters. As an important aside, note that all matrices  $D_{x}$, $D_{y}$, $D_{z}$, $\Wdepth$, $\Wh$, and $\WL$, are  sparse and  can therefore be saved using a MATLAB \sparse  format, with very limited demand on the  memory.

Now, in \eqref{OriginalP} both $\WL$ and $\bft(\bfm)$ depend on the model parameters $\bfm$. Thus,  the objective function \eqref{OriginalP} is nonlinear with respect to $\bfm$. We use a simple iterative strategy to convert $P^{(\alpha, \lambda)}(\bfm)$ to a linear form, in which to linearize the cross-gradient constraint, the first order Taylor expansion is used \cite{GM:03,GM:04,TL:06,Gallardo:07,FG:09}. 
First, suppose that the superscript $\ell$ applied to any variable indicates the value of that variable at iteration $\ell$, so that $\bfm^{(\ell)}$ is the estimate of the model parameters at iteration $ \ell$. Then, we suppose that $\bfm^{(1)} =\bfma$ and rewrite the objective function \eqref{OriginalP} as 
\begin{align}\label{ConvertedP}
P^{(\alpha,\lambda)}(\bfm)=\| \Wd (\bfdo-G \bfm)\|_2^2 + \alpha^{2}~\| W^{(\ell-1)}D(\bfm-\bfm^{(\ell-1)}) \|_2^2  \\ +  \lambda^{2}~\| \bft^{(\ell-1)}+\bfB^{(\ell-1)}(\bfm -\bfm^{(\ell-1)}) \|_2^2 \nonumber, \quad  \ell=2,3,...
\end{align}
Note that $W^{(\ell-1)}$ indicates $W$ estimated at iteration $ \ell-1$ through the nonlinear definition for $\WL$ as given in \eqref{WL1} or \eqref{WL2}. 
Here, $\bft^{(\ell-1)}$ and $\bfB^{(\ell-1)}=(\nabla_{\bfm}\bft^{(\ell-1)})$ are the cross-gradient and the Jacobian matrix of the discrete approximation for the cross-gradient function, respectively, evaluated at $\bfm^{(\ell-1)}$, consistent with the linear Taylor expansion for $\bft$ around $\bfm^{(\ell-1)}$. The formulae used are given in Appendix~\ref{App1}. Taking $\nabla_{\bfm}P^{(\alpha,\lambda)}{(\bfm)}=\mathbf{0}$  defines the update $\bfm^{(\ell)}$ as the solution of 
\begin{align}
-G^{T} \Wd^{T} \Wd (\bfdo-G \bfm) + \alpha^{2} D^{T}(W^{(\ell-1)})^TW^{(\ell-1)} D (\bfm-\bfm^{(\ell-1)}) + \\
\lambda^{2}~\left(  (\bfB^{(\ell-1)})^T \left\lbrace  \bft^{(\ell-1)}+\bfB^{(\ell-1)}(\bfm -\bfm^{(\ell-1)})\right\rbrace  \right)  = 0. \nonumber
\end{align}
Then, after some  algebraic manipulation, the desired update for $\bfm$ is given by
\begin{align}\label{updatem}
&\bfm^{(\ell)} = \bfm^{(\ell-1)} + \Delta \bfm^{(\ell-1)},
\end{align} 
where $\Delta \bfm^{(\ell-1)}$ is the solution of the equation 
\begin{align}\label{dmk}
\left(  G^{T} \Wd^{T} \Wd G  + \alpha^{2} D^{T} (W^{(\ell-1)})^TW^{(\ell-1)} D  + \lambda^{2}~ (\bfB^{(\ell-1)})^T \bfB^{(\ell-1)} \right)  \Delta \bfm^{(\ell-1)} =  \nonumber \\
\quad \left(G^{T} \Wd^{T} \Wd (\bfdo - G \bfm^{(\ell-1)}) - \lambda^{2}~ (\bfB^{(\ell-1)})^T \bft^{(\ell-1)} \right).
\end{align}
Equivalently, defining  
\begin{align}\label{Ek}
E^{(\ell)}=\left(  G^{T} \Wd^{T} \Wd G  + \alpha^{2} D^{T} (W^{(\ell-1)})^TW^{(\ell-1)} D  + \lambda^{2}~ (\bfB^{(\ell-1)})^T \bfB^{(\ell-1)} \right), 
\end{align} 
and 
\begin{align}\label{fk}
\bff^{(\ell)}=\left( G^{T} \Wd^{T} \Wd (\bfdo - G \bfm^{(\ell-1)}) - \lambda^{2}~ (\bfB^{(\ell-1)})^T \bft^{(\ell-1)} \right),  
\end{align}
then $ \Delta \bfm^{(\ell-1)}$ given by \eqref{dmk} solves the linear system $E^{(\ell)} \Delta \bfm^{(\ell-1)}=\bff^{(\ell)}$.  Numerically the CG algorithm can be used to solve \eqref{dmk} and a line search can also be included in the update \eqref{updatem}, replacing $\Delta \bfm^{(\ell-1)}$ by $\gamma^{(\ell-1)}\Delta \bfm^{(\ell-1)}$ where $0<\gamma^{(\ell-1)}$ is chosen to speed convergence, see for example Fournier \& Oldenburg \shortcite{FoOl:19}. We should note that at each iteration of the algorithm lower and upper bounds on density and susceptibility are imposed. During the inversion process if an estimated physical property falls outside the specified bounds, it will be returned back to the nearest bound. Further, to test the convergence of the solution at each iteration  $\ell$, we calculate   a $\chi^2$ measure for the respective data fit term at each iteration, 
\begin{align}\label{Chi1}
(\chii^{2})^{(\ell)}=\|  \Wdi (\bfdi-\Gi \bfmi^{(\ell)}) \|_{2}^{2}, \quad \textrm{i}=1, 2.
\end{align}
The iteration will terminate at convergence only when  $(\chii^{2})^{(\ell)} \leq  m + \sqrt{2m}$, for both $\textrm{i}=1$, and $\textrm{i}=2$. Otherwise, the iteration is allowed to proceed to a maximum number of iterations $\Kmax$. The steps of the joint inversion algorithm are summarized in Algorithm~\ref{Jointalgorithm}. 

In \eqref{OriginalP} the  important parameters $\alpha$ and $\lambda$ are the regularization parameters which give relative weights to the stabilizer and the cross-gradient term,  respectively. To be more precise, we define  $\alpha$ as $\blkdiag\left( \alphag I_{n},  \alpham I_{n} \right) \in \Rmn{2n}{2n}$, where $\alphag$ and $\alpham$ are the relative weights for the gravity and magnetic terms,  respectively. We should note that, although $\alpha$ is a diagonal matrix and can be used inside the stabilizer, we prefer to put $\alpha$ outside in order for the formulation to be consistent with the  conventional Tikhonov objective functional.  Weighting parameters $\alphag$, $\alpham$, and $\lambda$ have an important effect on the estimated solution. Thus, they need to be determined carefully. But  application of an automatic parameter-choice method for determining $\alphag$, $\alpham$, and $\lambda$ is difficult, or potentially impossible, and is outside the scope of this current study. Therefore we adopt a simple but practical strategy for determining  suitable values of these parameters. Previous investigations have demonstrated that it is efficient if the inversion starts with a large regularization parameter \cite{Far:2004,VAR:2014a}. We follow that strategy here and start the inversion with  large $\alphag$ and $\alpham$. In subsequent iterations  the parameters are reduced slowly dependent on  parameters $\qg$ and $\qm$, respectively, using $\alphag^{(\ell)}=\alphag^{(\ell-1)}\qg$ and $\alpham^{(\ell)}=\alpham^{(\ell-1)}\qm$, where $\qg$ and $\qm$ are small numbers, $0 \ll \qg, \qm  <1$. The process continues until the predicted data of one of the reconstructed models satisfies the observed data at the noise level. For that data set, the relevant parameter is then kept fixed during the following iterations. The parameter $\lambda$  is held fixed  in the implementation, although it is quite feasible that it is also iteration dependent.  The  amount of structural similarity obtained through  the joint inversion algorithm can be adjusted using different choices of $\lambda$. The impact of selecting the parameters $\alphag$, $\alpham$, (equivalently $\qg$ and $\qm$), and $\lambda$ is demonstrated in Section~\ref{Synthetic}.

\begin{algorithm}
\caption{Generalized $L_{p}$-norm joint inversion of gravity and magnetic data.}\label{Jointalgorithm}
\begin{algorithmic}[1]
\REQUIRE $\bfdo$, $\bfma$, $G$, $\Wd$, $\Whg$, $\Whm$, $D_{x}$, $D_{y}$, $D_{z}$, $\epsilon$, $\rho_{\mathrm{min}}$, $\rho_{\mathrm{max}}$, $\kappa_{\mathrm{min}}$ , $\kappa_{\mathrm{max}}$, $\Kmax$, $\betag$, $\betam$, $\alphag^{(1)}$, $\alpham^{(1)}$, $\qg$, $\qm$ and $\lambda$.
\STATE Calculate $\Wdepthg$ and $\Wdepthm$ as determined by $\betag$ and $\betam$, respectively.
\STATE Set $\Wg =\Wdepthg \Whg$ and  $\Wm =\Wdepthm \Whm$. 
\STATE If $D$ is the identity form $W=\blkdiag\left(\Wg,\Wm\right)$. Otherwise set $W=\blkdiag\left(\Wg,\Wg,\Wg,\Wm,\Wm,\Wm\right)$.
\STATE Initialize $\bfm^{(1)}=\bfma$, $\WLg^{(1)}=I$, $\WLm^{(1)}=I$, and $ \ell=1$. 
\WHILE {Not converged, noise level not satisfied, and $ \ell<\Kmax$} 
\STATE $ \ell= \ell+1$.
\STATE Compute $\bft^{(\ell-1)}$ and $\bfB^{(\ell-1)}$,    {as given in Appendix~\ref{App1}.} 
\STATE Use CG to solve  $E^{(\ell)}\Delta \bfm^{(\ell-1)}=\bff^{(\ell)}$ for $\bfm^{(\ell-1)}$, defined by \eqref{Ek} and \eqref{fk}.
\STATE  Update $\bfm^{(\ell)} = \bfm^{(\ell-1)} + \Delta \bfm^{(\ell-1)}$.
\STATE Impose constraints on $\bfm^{(\ell)}$ to force $\rho_{\mathrm{min}}\le \bfmg^{(\ell)} \le \rho_{\mathrm{max}}$ and $\kappa_{\mathrm{min}}\le \bfmm^{(\ell)} \le \kappa_{\mathrm{max}}$.
\STATE Test convergence criteria, \eqref{Chi1}, for $\chig^{2}$ and $\chim^{2}$. Exit loop if both satisfied.
\STATE Set $\alphag^{(\ell)}=\alphag^{(\ell-1)}\qg$ and $\alpham^{(\ell)}=\alpham^{(\ell-1)} \qm$. Update $\alpha^{(\ell)}$.
\STATE Determine $\WLg^{(\ell)}$ and $\WLm^{(\ell)}$, dependent on $D$ and $p$, equations \eqref{WL1} or \eqref{WL2}.
\STATE Set $W^{(\ell)}=\blkdiag\left(\Wg^{(\ell)},\Wm^{(\ell)}\right)$ when $D$ is the identity. Otherwise set  $W^{(\ell)}=\blkdiag\left(\Wg^{(\ell)},\Wg^{(\ell)},\Wg^{(\ell)},\Wm^{(\ell)},\Wm^{(\ell)},\Wm^{(\ell)}\right)$
\ENDWHILE
\ENSURE Solution $\rho=\bfmg^{(\ell)}$, $\kappa=\bfmm^{(\ell)}$, $\mathrm{IT}= \ell$.
\end{algorithmic}
\end{algorithm}

\section{Simulations}\label{Synthetic}
The validity of the presented joint inversion algorithm is evaluated for synthetic examples and the results are compared with those of the separate gravity and magnetic inversions. We select a model that consists of  two dikes, one is a small vertical dike and the other is a larger dipping dike. In the first study, we suppose that the targets have both a density contrast and a susceptibility distribution. The density contrast and the susceptibility of the targets are selected as $\rho=0.6$~gr~cm$^{-3}$ and $\kappa=0.06$ (SI unit), respectively, embedded in a homogeneous non-susceptible background. For the second study, we assume that the dipping dike has a  density and susceptibility distribution, but that the vertical dike   is not magnetic.  For all inversions, the bound constraints $0=\rho_{\mathrm{min}}\leq \bfmg \leq \rho_{\mathrm{max}}=0.6$, gr~cm$^{-3}$, and $0=\kappa_{\mathrm{min}}\leq \bfmm \leq \kappa_{\mathrm{max}}=0.06$, SI unit, are imposed. To generate the total field anomaly, the intensity of the geomagnetic field, the inclination, and the declination are selected as $50000$ nT, $45^{\circ}$ and $45^{\circ}$, respectively.  To simulate realistic noise-contaminated   observed data, we suppose that Gaussian noise with zero mean and standard deviations $(\tau_{1}~| (\bfdei)_{j}|+\tau_{2}~ \text{max}(| \bfdei |))$, $j=1...n$, are added  to each exact true   measurement $j$, $(\bfdei)_j$. 
The parameters pairs $(\tau_{1}, \tau_{2})$ are selected as $(0.02, 0.012)$ and $(0.02, 0.01)$ for gravity and magnetic data, respectively. These standard deviations are selected such that the signal to noise ratios (SNR), as given by
\begin{align}\label{snr}
\SNRi=20~\text{log}_{10}(\frac{\| \bfdei \|_{2}}{\| \bfdi-\bfdei  \|_{2}}), \quad \textrm{i}=1, 2,
\end{align}
become nearly the same.  The resulting SNRs are indicated in the captions of the respective figures associated with the results for the given data set. In all simulations we use $\betag=0.8$ and $\betam=1.4$ in $\Wdepthg$ and $\Wdepthm$, respectively. The maximum number of iterations of the algorithm is  selected as $\Kmax=100$ for all inversions. Further, in the following, we present the results for two different stabilizers;  the L$_{1}$-norm of the model parameters imposed using $D=I$ and $p=1$ in \eqref{WL1}, and the minimum structure stabilizer, imposed using $p=2$ and $D \neq I$ in \eqref{WL2}. This then demonstrates the validity of the algorithm for both sparse and smooth reconstructions of the   subsurface targets. Focusing parameter $\epsilon$ is held fixed in all inversions,  $\epsilon^{2}=1e^{-9}$. A summary of the parameters chosen for the simulations discussed in Sections~\ref{model1}-\ref{model2} are provided in Table~\ref{table:parameters}.

\begin{table}
\begin{center}
\begin{tabular}{ccccccccccccc}\hline
Model&Case&Figure&$p$&$D$&$\alphag^{(1)}$&$\qg$&$\alpham^{(1)}$&$ \qm$&$\lambda$&$\Kmax$&$\Wh$&Results\\ \hline
$1$&$1$&$1$&$1$&$I$&$20,000$& $0.9$&$50,000$&$0.95$&$0$&$100$ &-&\ref{fig3}\\
$1$&$2$&$1$&$1$&$I$&$20,000$& $0.9$&$50,000$&$0.95$&$10^6$&$100$ &-&\ref{fig6}\\
$1$&$2$&$1$&$1$&$I$&$20,000$& $0.9$&$50,000$&$0.95$&$10^{12}$&$100$ &-&\ref{fig9} \\
$1$&$3$&$1$&$1$&$I$&$20,000$& $0.9$&$50,000$&$0.95$&$10^6$&$100$ &Yes&\ref{fig12} \\
$1$&$4$&$1$&$2$&$D$&$200,000$& $0.9$&$500,000$&$0.95$&$10^7$&$100$ &-&\ref{fig13} \\ \hline
$2$&$1$&$16$&$1$&$I$&$20,000$& $0.9$&$50,000$&$0.95$&$10^6$&$100$ &-&\ref{fig18}\\
$2$&$2$&$16$&$2$&$D$&$200,000$& $0.9$&$500,000$&$0.95$&$10^7$&$100$ &-&\ref{fig19}\\ \hline
\end{tabular}
\end{center}
\caption{Summary of the parameters chosen for each of the synthetic simulations that are discussed in Sections~\ref{model1} and \ref{model2}. The true cross section of the models are provided in the figure indicated by the number in the column with heading Figure. The obtained cross sections by the inversion are illustrated in the figure with number in the column with heading Results. \label{table:parameters}}
\end{table}

\subsection{Model study 1}\label{model1}
\begin{figure*}
\subfigure{\label{fig1a}\includegraphics[width=0.49\textwidth]{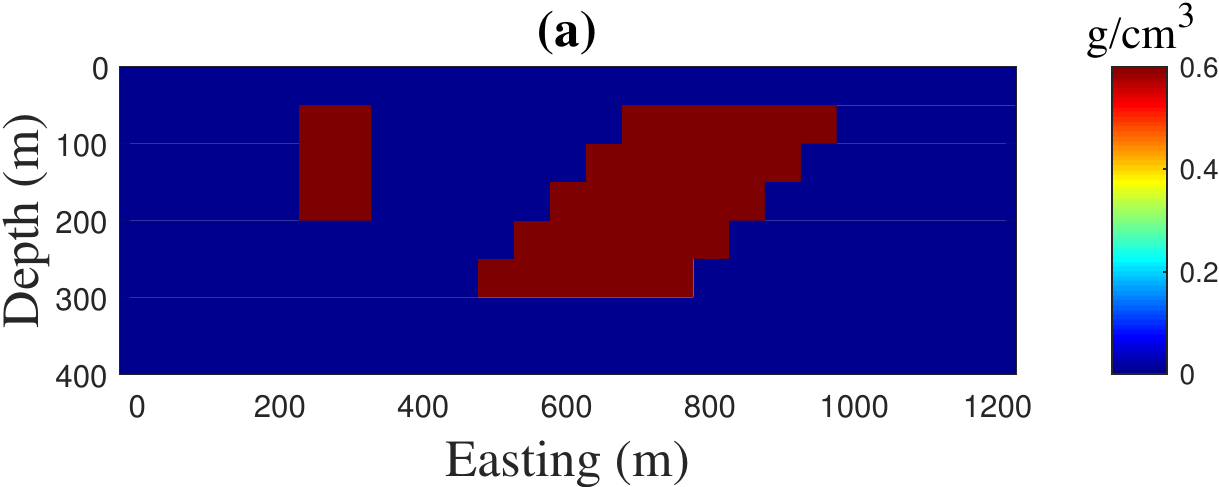}}
\subfigure{\label{fig1b}\includegraphics[width=0.49\textwidth]{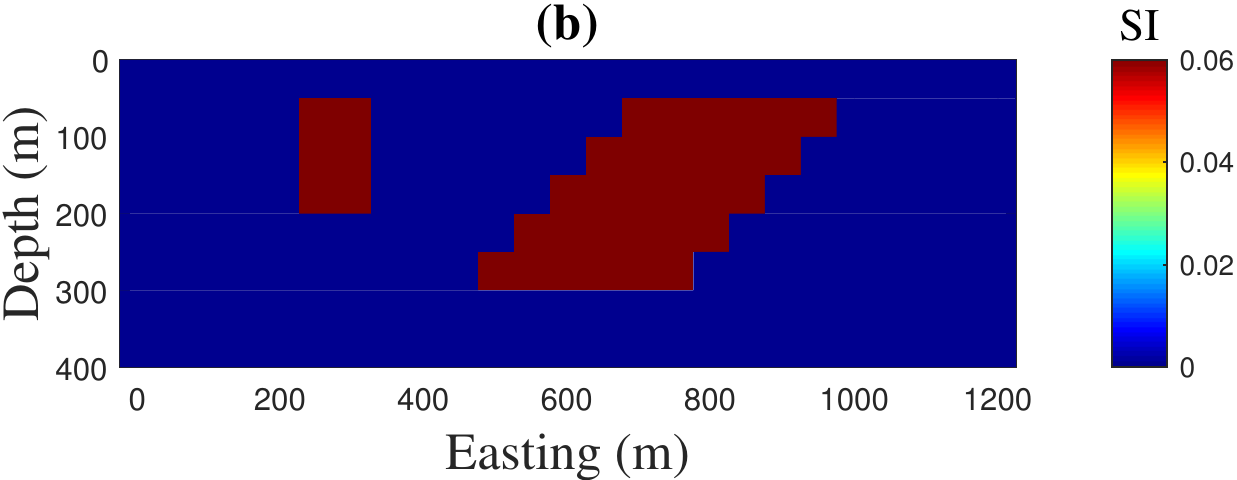}}
\caption {Cross-section of the synthetic model that consists of a  vertical and a dipping dike. (a) Density distribution; (b) Susceptibility distribution.}\label{fig1}
\end{figure*}

\begin{figure*}
\subfigure{\label{fig2a}\includegraphics[width=0.47\textwidth]{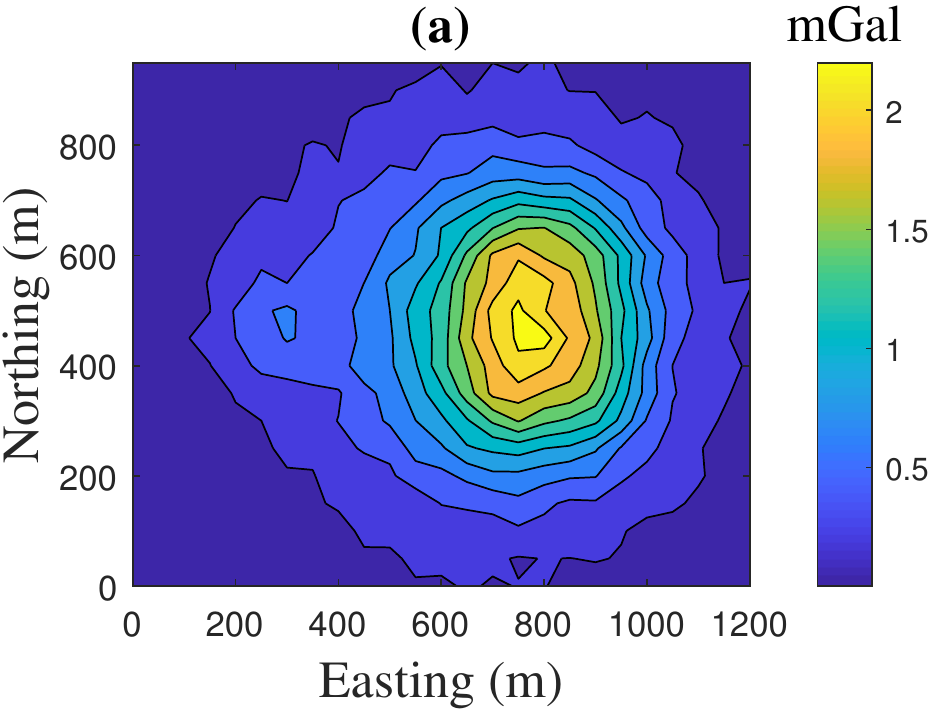}}
\subfigure{\label{fig2b}\includegraphics[width=0.48\textwidth]{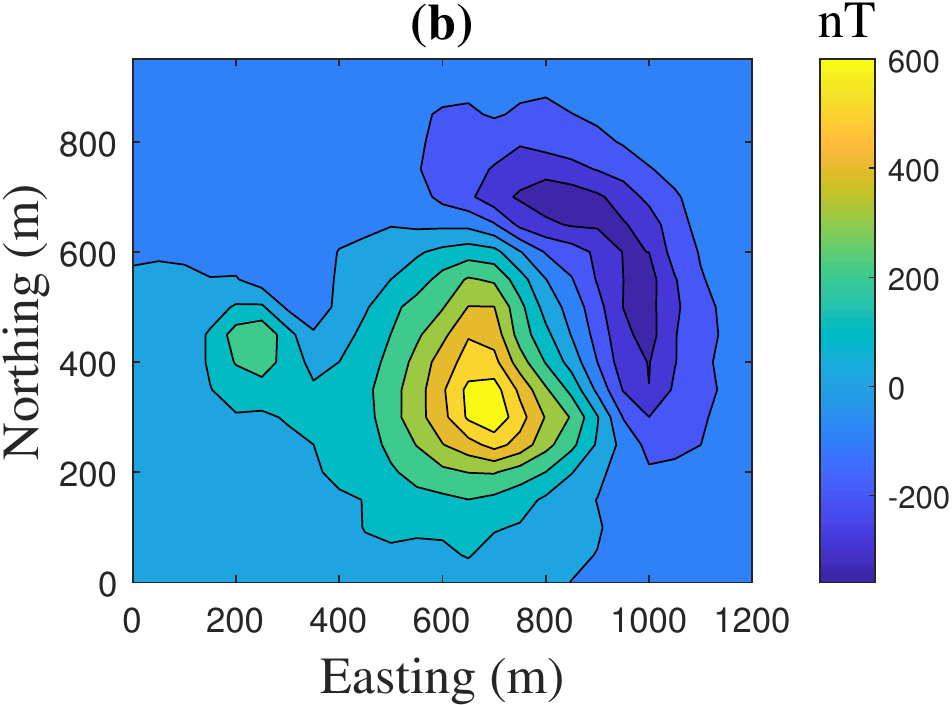}}
\caption {Noise contaminated anomaly produced by the model shown in Fig.~\ref{fig1}. (a) Vertical component of gravity; (b) Total magnetic field. The SNR for gravity and magnetic data, respectively, are $25.1778$ and $25.1558$.}\label{fig2}
\end{figure*}

The first model is illustrated in Fig.~\ref{fig1}. The top depths of the bodies are located at $50$~m. The dipping dike is extended to $300$~m, but the maximum depth of the small vertical dike is $200$~m. The data on the surface are generated on a grid with $25 \times 20 =500$ points and grid spacing $50$~m. The noise-contaminated gravity and magnetic data are illustrated in Figs.~\ref{fig2a} and \ref{fig2b}, respectively. To perform the inversion, the subsurface volume is discretized into   $4000$ prisms of sizes $50$~m in each dimension. The initial regularization parameters are selected as  $\alphag^{(1)}=20,000$ and $\alpham^{(1)}=50,000$. In Vatankhah et al. \shortcite{VLRHB:19b} we demonstrated that the gravity and magnetic sensitivity matrices, $\Gg$ and $\Gm$, have different spectral properties and, therefore, the regularization parameter should be much larger for the inversion of magnetic data as compared to that used for the inversion of gravity data.  Thus, it is appropriate to  control the speed of convergence for each model with parameters $\qg=0.9$ and $\qm=0.95$, that are different.

\subsubsection{Case 1: separate L$_{1}$-norm inversion}\label{Case1}
We first implement the algorithm using the L$_{1}$-norm stabilizer but without using the cross-gradient constraint. This is easy to do by selecting $\lambda=0$.  Hence the algorithm proceeds exactly as in Algorithm~\ref{Jointalgorithm}, with the same termination criteria, but without the cross-gradient term.  The inversion  is initiated using $\bfma=\mathbf{0}$. After $\mathrm{IT}=61$ iterations the convergence criteria, $\chig^2$ and $\chim^2$, are satisfied and the inversion terminates. In this simulation the $\chim^2$ termination is reached at iteration $43$.  The susceptibility model is recovered more quickly than the density model, which requires additional iterations until both termination criteria are satisfied. 

\begin{figure*}
\subfigure{\label{fig3a}\includegraphics[width=0.49\textwidth]{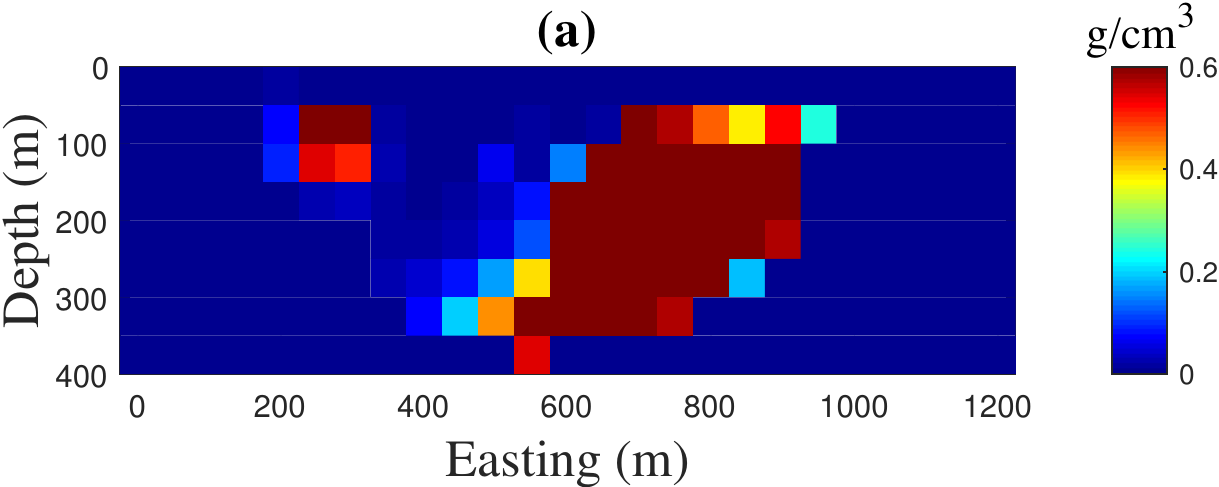}}
\subfigure{\label{fig3b}\includegraphics[width=0.49\textwidth]{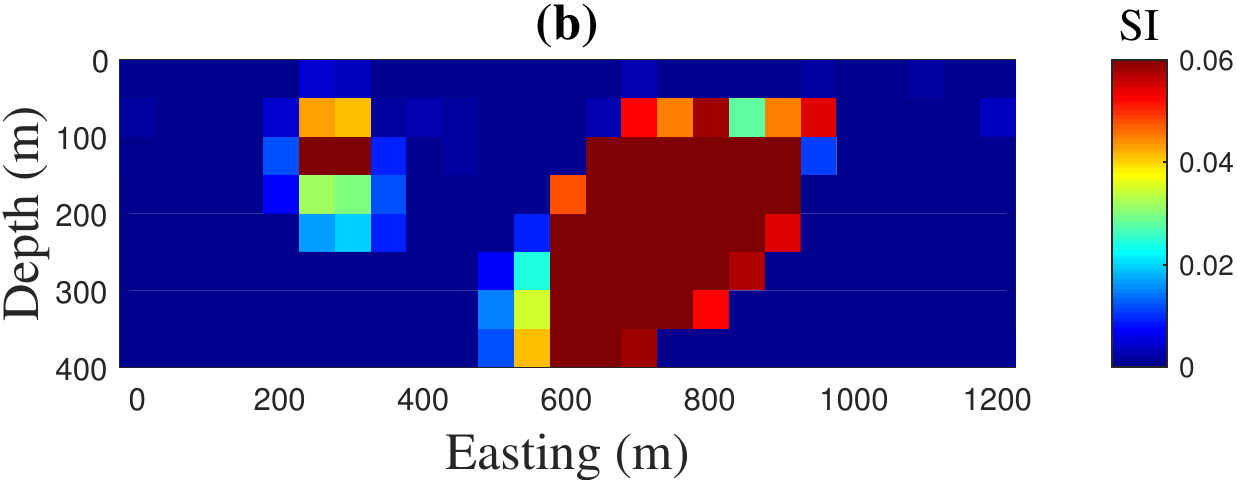}}
\caption {Cross-section of the reconstructed model using individual inversions, Case~$1$. (a) Density distribution; (b) Susceptibility distribution.}\label{fig3}
\end{figure*}

\begin{figure*}
\subfigure{\label{fig4a}\includegraphics[width=0.47\textwidth]{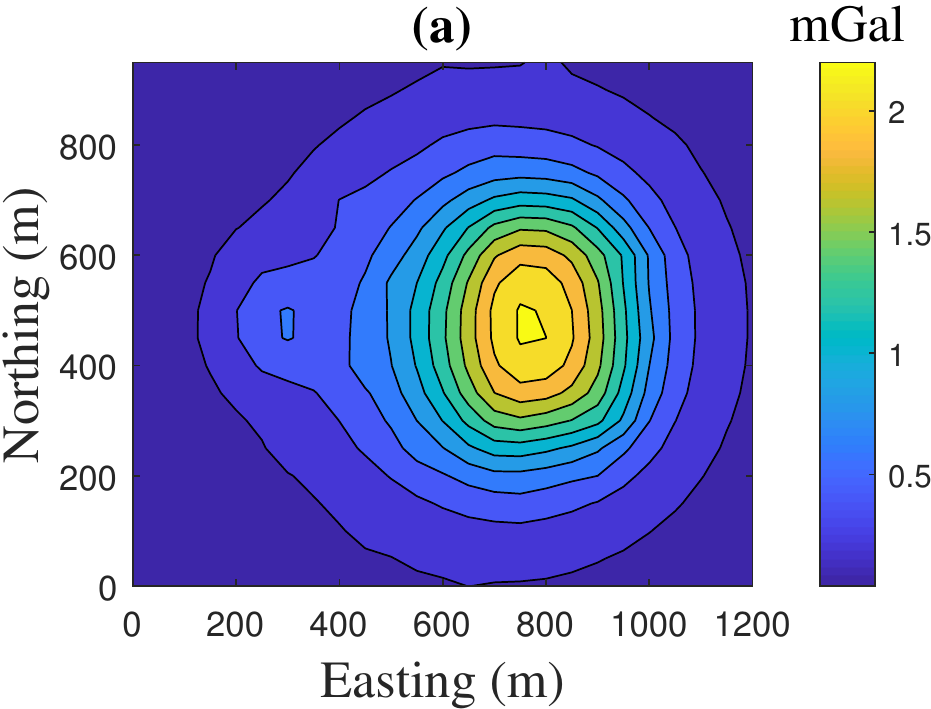}}
\subfigure{\label{fig4b}\includegraphics[width=0.48\textwidth]{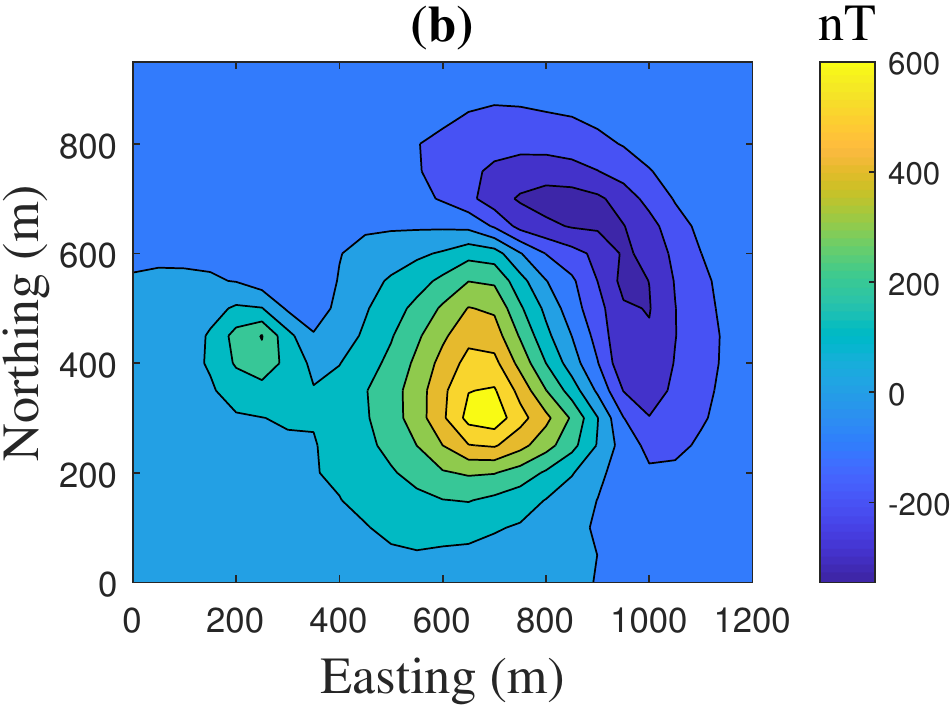}}
\caption {The data produced by the models shown in Fig.~\ref{fig3}, Case~$1$. (a) Vertical component of gravity; (b) Total magnetic field.}\label{fig4}
\end{figure*}

\begin{figure*}
\subfigure{\label{fig5a}\includegraphics[width=0.49\textwidth]{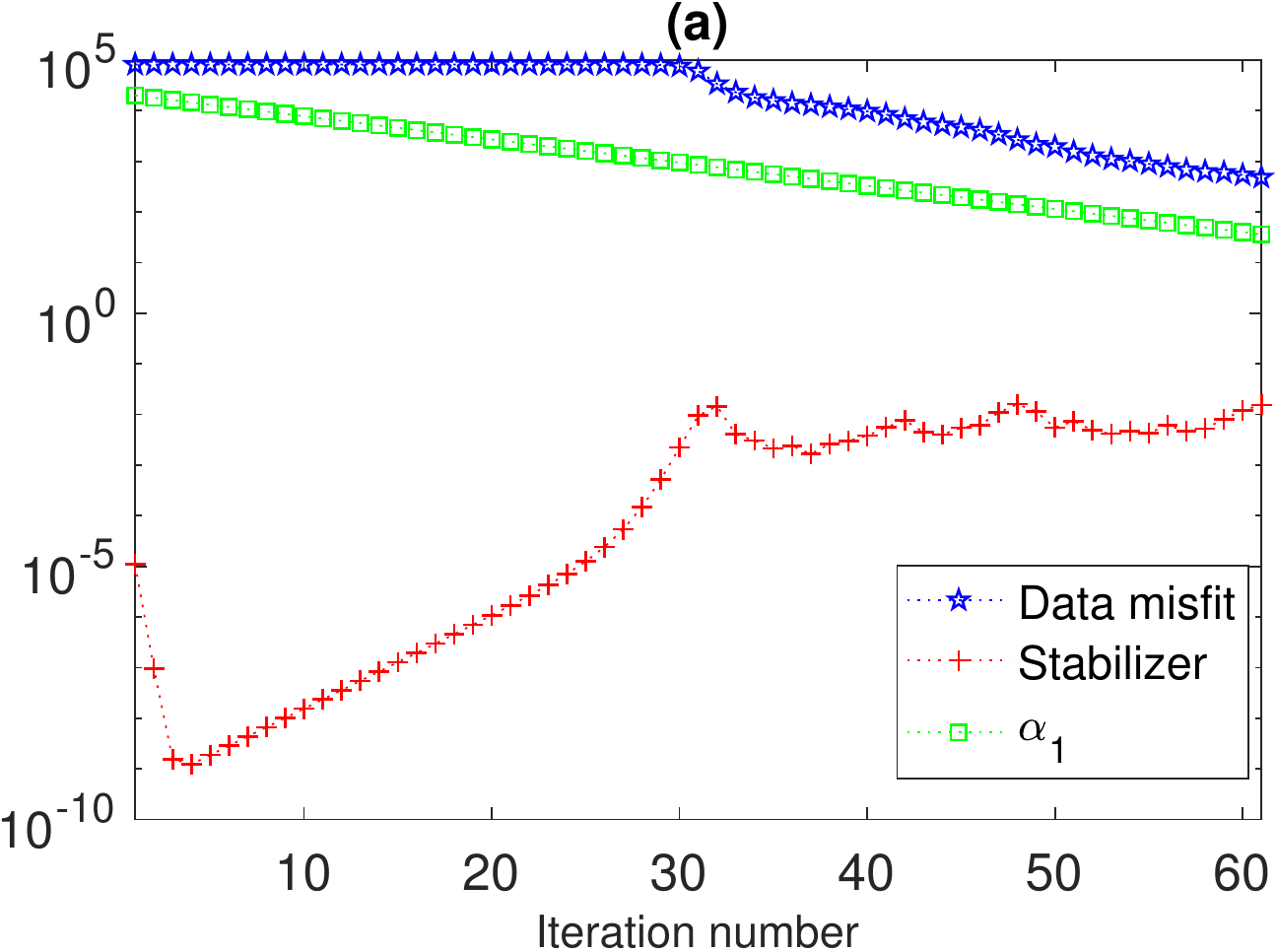}}
\subfigure{\label{fig5b}\includegraphics[width=0.49\textwidth]{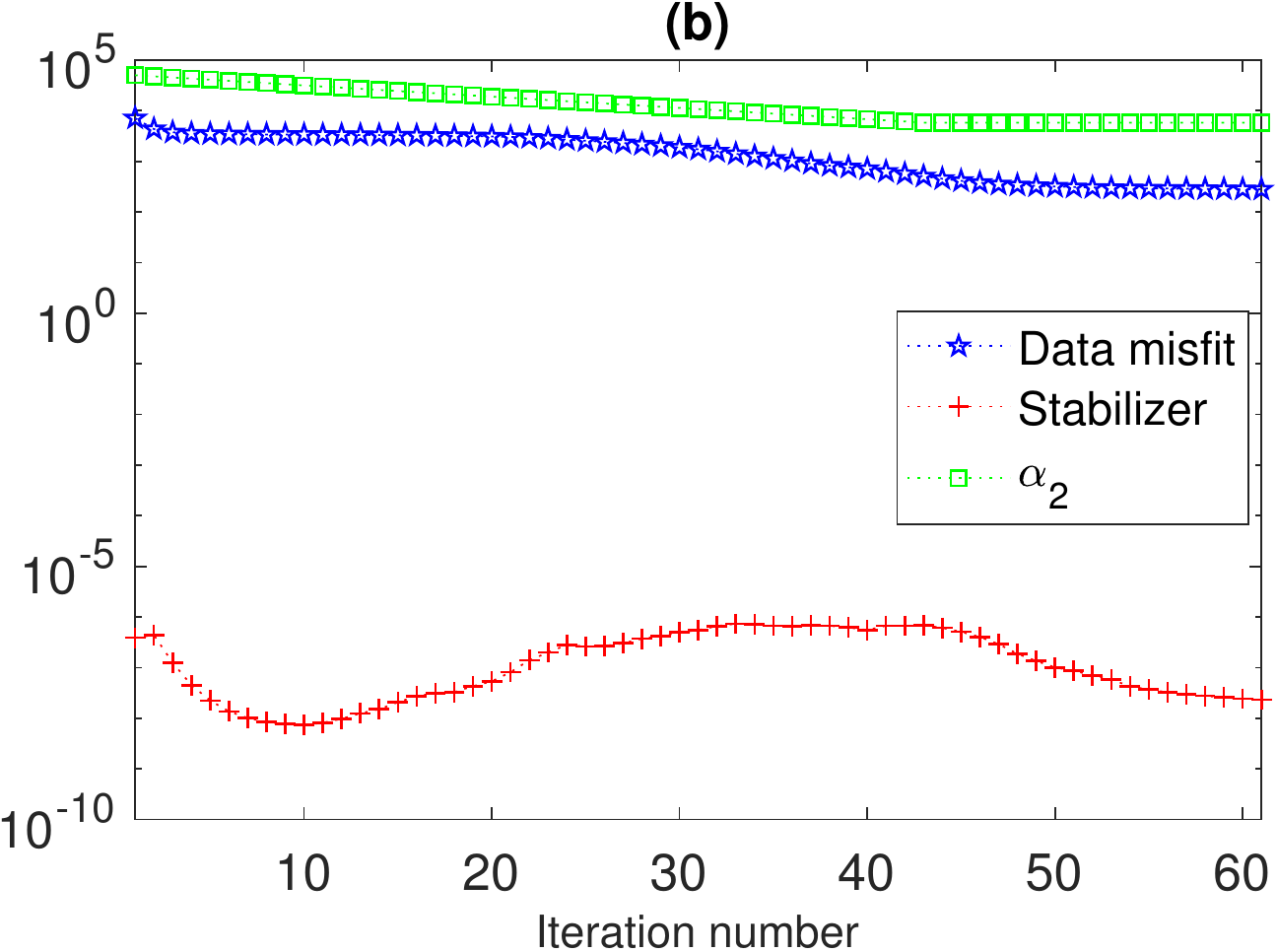}}
\caption {The progression of the data misfit, the regularization term, and the regularization parameter, with iteration  $\ell$, for the models shown in Fig.~\ref{fig3}, Model~$1$ and Case~$1$. (a) Gravity inversion; (b) Magnetic inversion.}\label{fig5}
\end{figure*}

The reconstructed density and susceptibility models are illustrated in Figs.~\ref{fig3a} and \ref{fig3b}, respectively. They are in good agreement with the original models, and are acceptable. The results are very close to those that can be obtained using another single L$_{1}$-norm algorithm, see Vatankhah et al. \shortcite{VLRHB:19b}, in which the unbiased predictive risk estimator is used as parameter-choice method and the singular value decomposition (SVD) is used computationally. It is clear that sharp and focused images of the subsurface are obtained, but the recovered susceptibility model shows more extension in depth for both targets. Further, in the density model, the extension of the vertical dike is underestimated. For comparison, we present the data produced by the models in Figs.~\ref{fig4a} and \ref{fig4b}, respectively. In addition, the the progression of the data misfit, the regularization term, and the regularization parameter, with iteration  $\ell$,  for both gravity and magnetic problems, are illustrated in Fig.~\ref{fig5}.

\subsubsection{Case 2: joint L$_{1}$-norm inversion with cross-gradient constraint}\label{Case2}
\begin{figure*}
\subfigure{\label{fig6a}\includegraphics[width=0.49\textwidth]{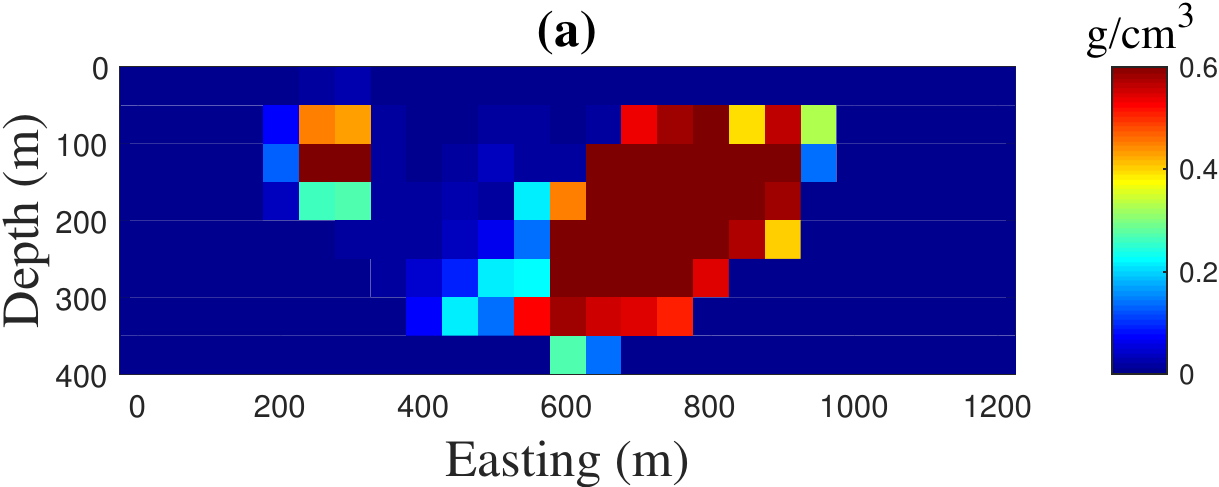}}
\subfigure{\label{fig6b}\includegraphics[width=0.49\textwidth]{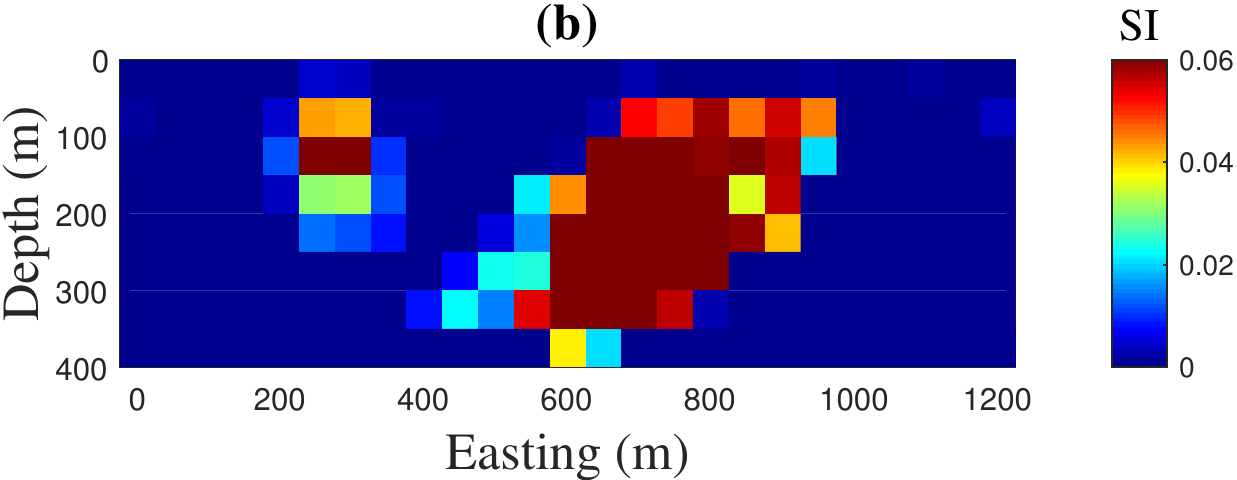}}
\caption {Reconstructed models using joint inversion with $\lambda=10^{6}$ and with the L$_{1}$-norm of the model parameters as the stabilizer, Model~$1$ and Case~$2$. (a) Density distribution; (b) Susceptibility distribution.}\label{fig6}
\end{figure*}

\begin{figure*}
\subfigure{\label{fig7a}\includegraphics[width=0.47\textwidth]{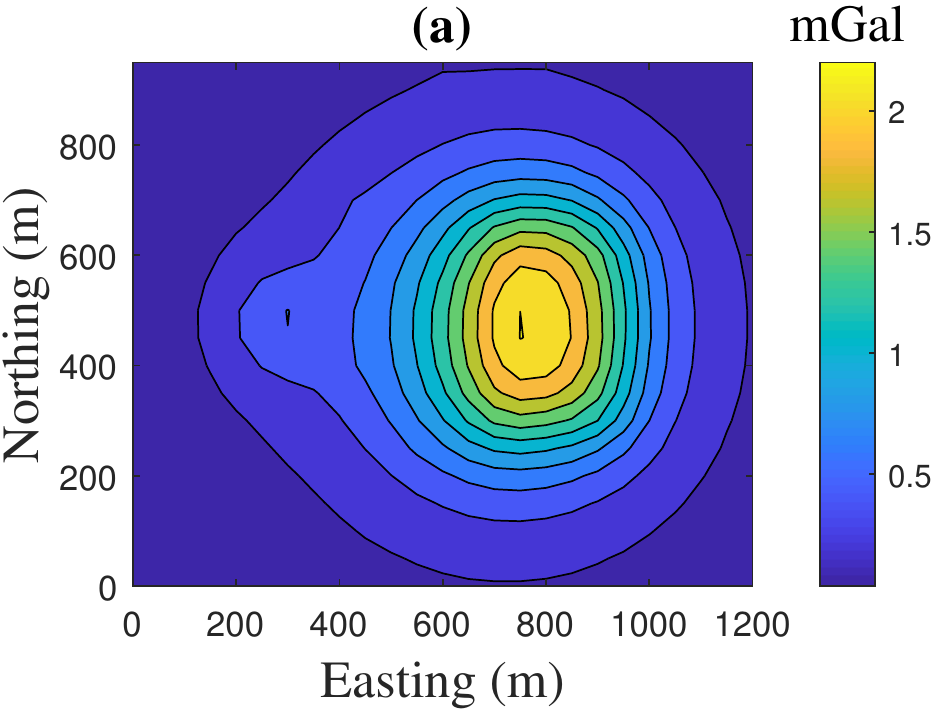}}
\subfigure{\label{fig7b}\includegraphics[width=0.48\textwidth]{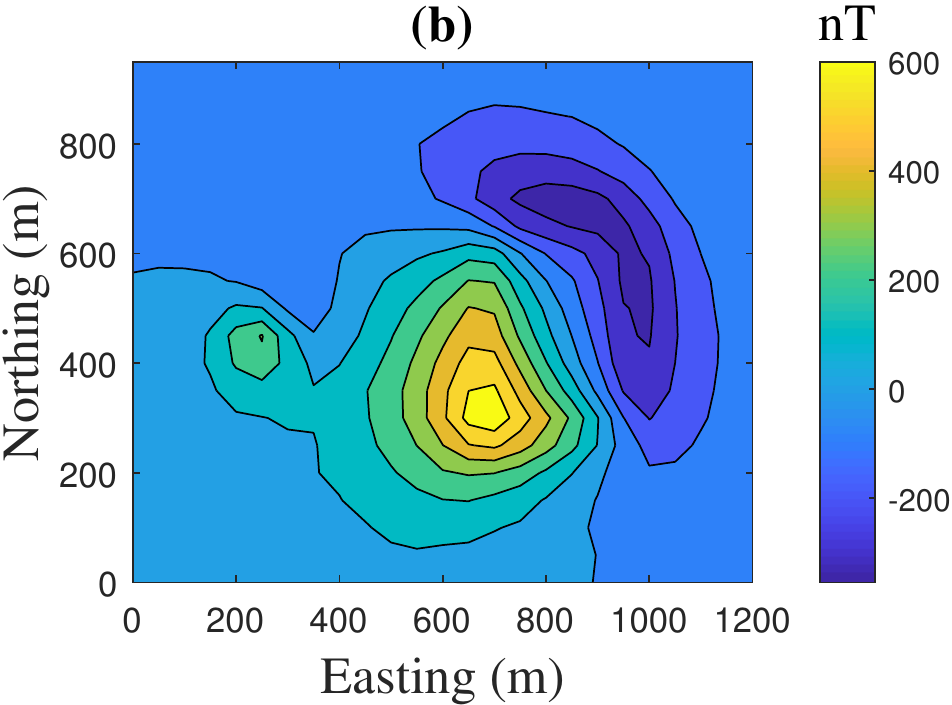}}
\caption {The data produced by the models shown in Fig.~\ref{fig6}, Model~$1$ and Case~$2$. (a) Vertical component of gravity; (b) Total magnetic field.}\label{fig7}
\end{figure*}

\begin{figure*}
\subfigure{\label{fig8a}\includegraphics[width=0.49\textwidth]{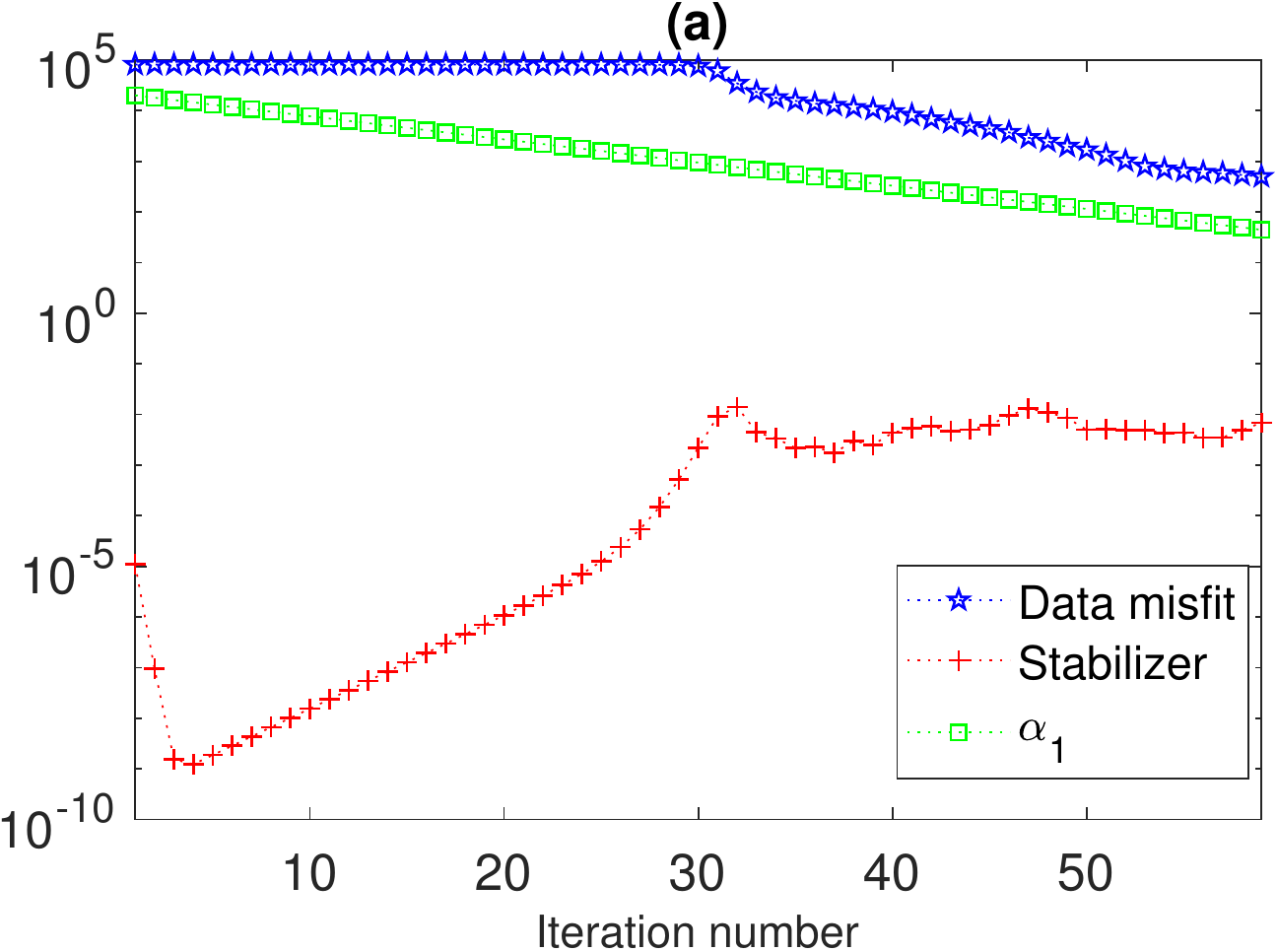}}
\subfigure{\label{fig8b}\includegraphics[width=0.49\textwidth]{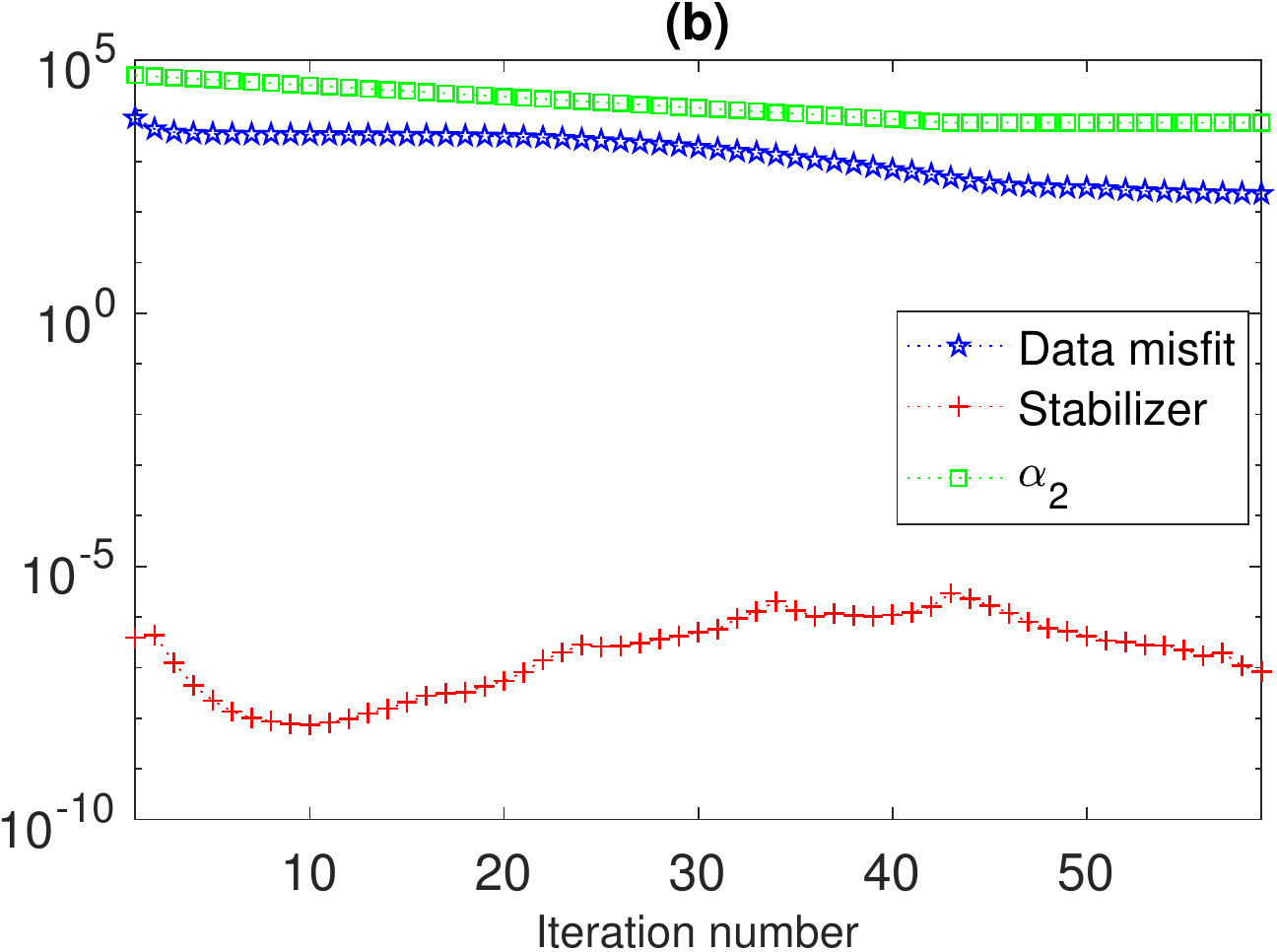}}
\caption {The progression of the data misfit, the regularization term, and the regularization parameter, with iteration  $\ell$, for the models shown in Fig.~\ref{fig6}, Model~$1$ and Case~$2$. (a) Gravity; and (b) Magnetic.}\label{fig8}
\end{figure*}

We now implement the inversion algorithm using the cross-gradient constraint with $\lambda=10^6$. This selection is based on an analysis of entries of matrix $\bfB$, which are very small. If  we want to give enough weight to the cross-gradient term, it is necessary to use a large value for $\lambda$.  We will also show, on the other hand, that  if $\lambda$ is too large the results can be unsatisfactory. All other parameters are selected  as for the simulation for Model~$1$ and Case~$1$, in Section~\ref{Case1}. The inversion terminates at iteration $\mathrm{IT}=59$, two iterations less than required for the presented individual inversions.  The results are presented in Figs.~\ref{fig6}, \ref{fig7}, and \ref{fig8}, respectively. Now, the reconstructed density and susceptibility models are quite similar, and are close to the original models. Indeed, the results indicate that the application of the cross-gradient function significantly improves the quality of the solutions. Both targets are in better agreement with the original models.

\begin{figure*}
\subfigure{\label{fig9a}\includegraphics[width=0.49\textwidth]{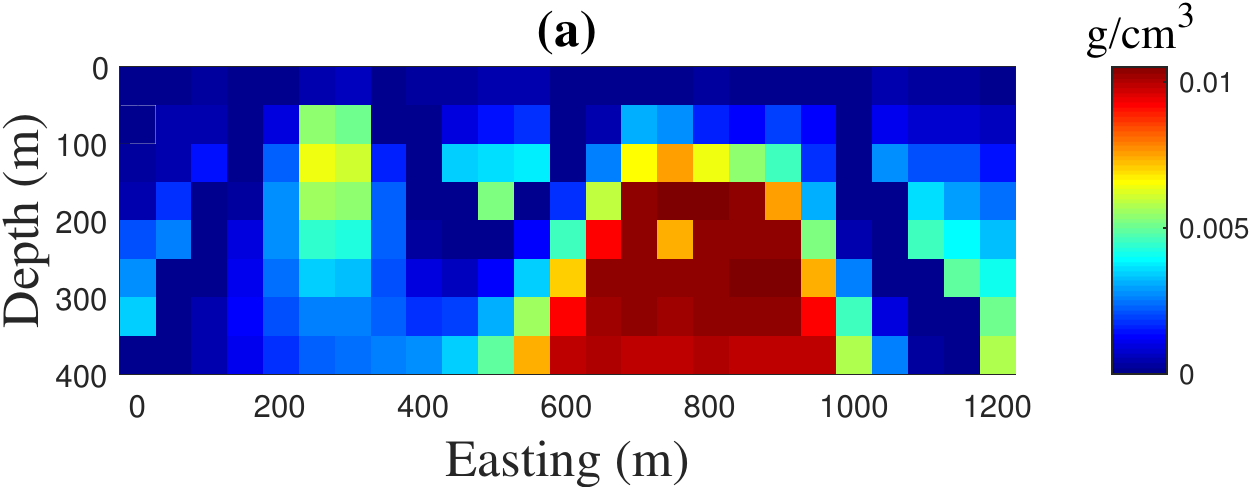}}
\subfigure{\label{fig9b}\includegraphics[width=0.49\textwidth]{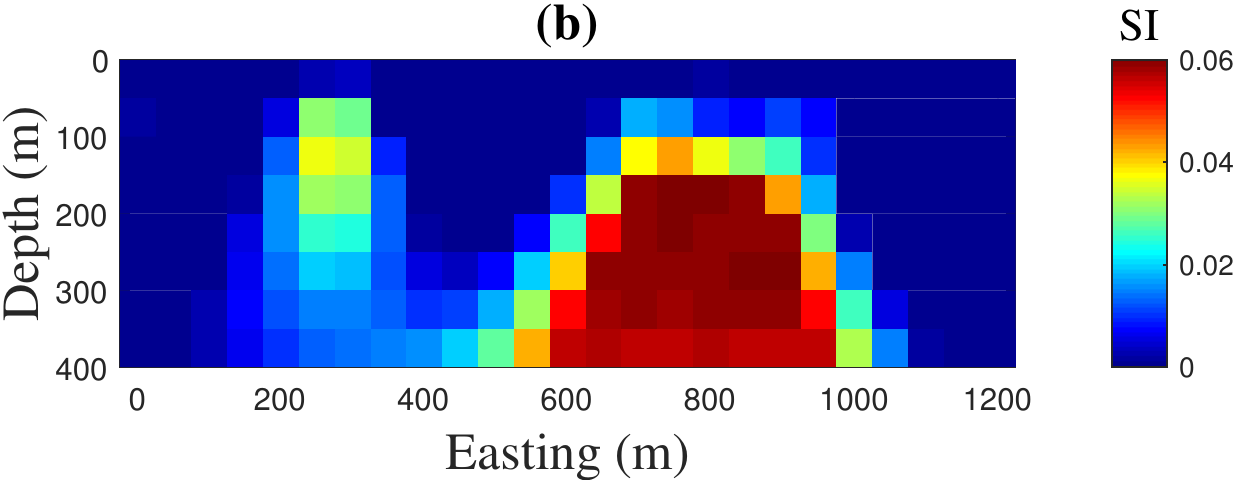}}
\caption {Reconstructed models using joint inversion with $\lambda=10^{12}$ and with the L$_{1}$-norm of the model parameters as the stabilizer, Model~$1$ and Case~$2$. (a) Density distribution; (b) Susceptibility distribution.}\label{fig9}
\end{figure*}

\begin{figure*}
\subfigure{\label{fig10a}\includegraphics[width=0.48\textwidth]{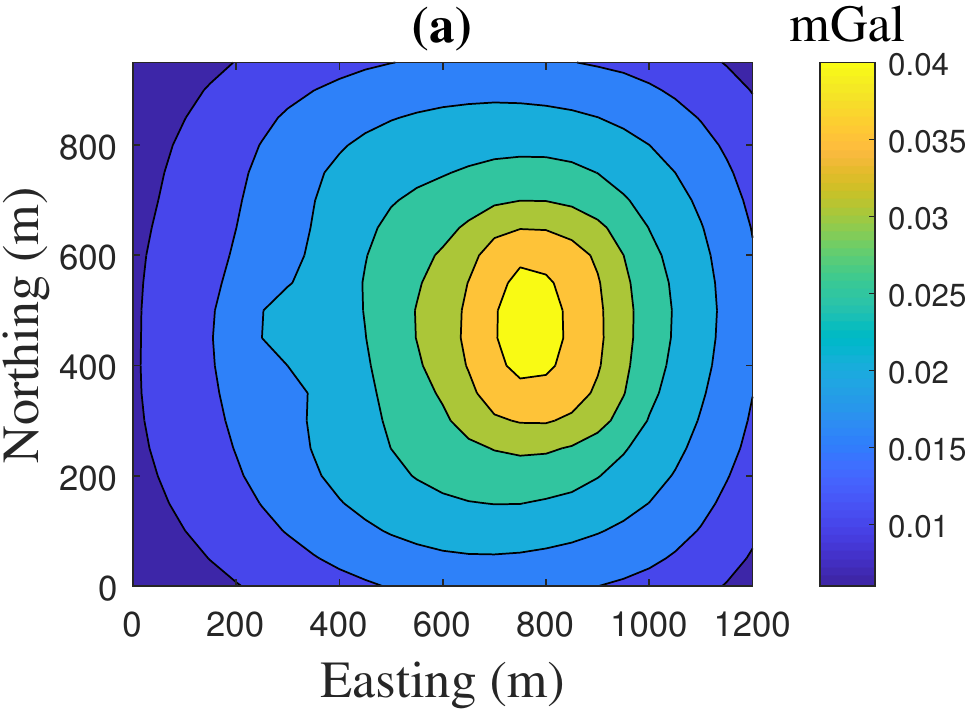}}
\subfigure{\label{fig10b}\includegraphics[width=0.48\textwidth]{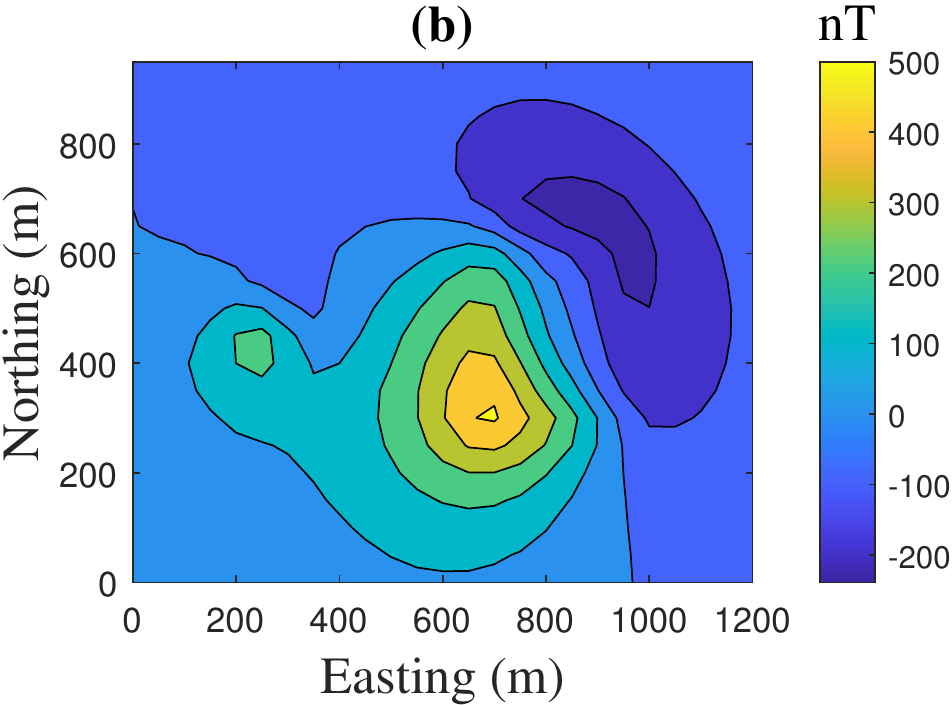}}
\caption {The data produced by the models shown in Fig.~\ref{fig9}, Model~$1$ and Case~$2$. (a) Vertical component of gravity; (b) Total magnetic field.}\label{fig10}
\end{figure*}

\begin{figure*}
\subfigure{\label{fig11a}\includegraphics[width=0.49\textwidth]{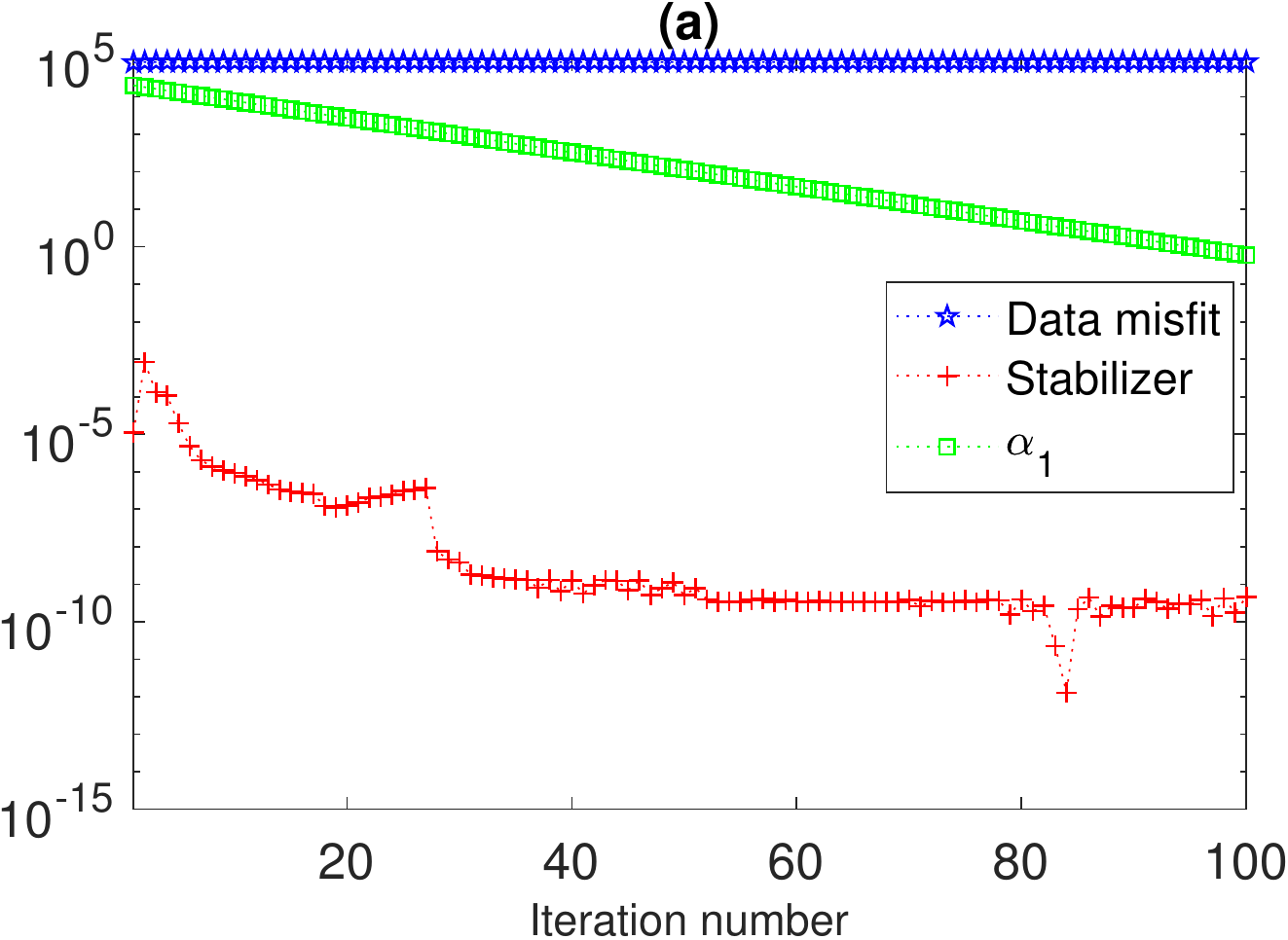}}
\subfigure{\label{fig11b}\includegraphics[width=0.49\textwidth]{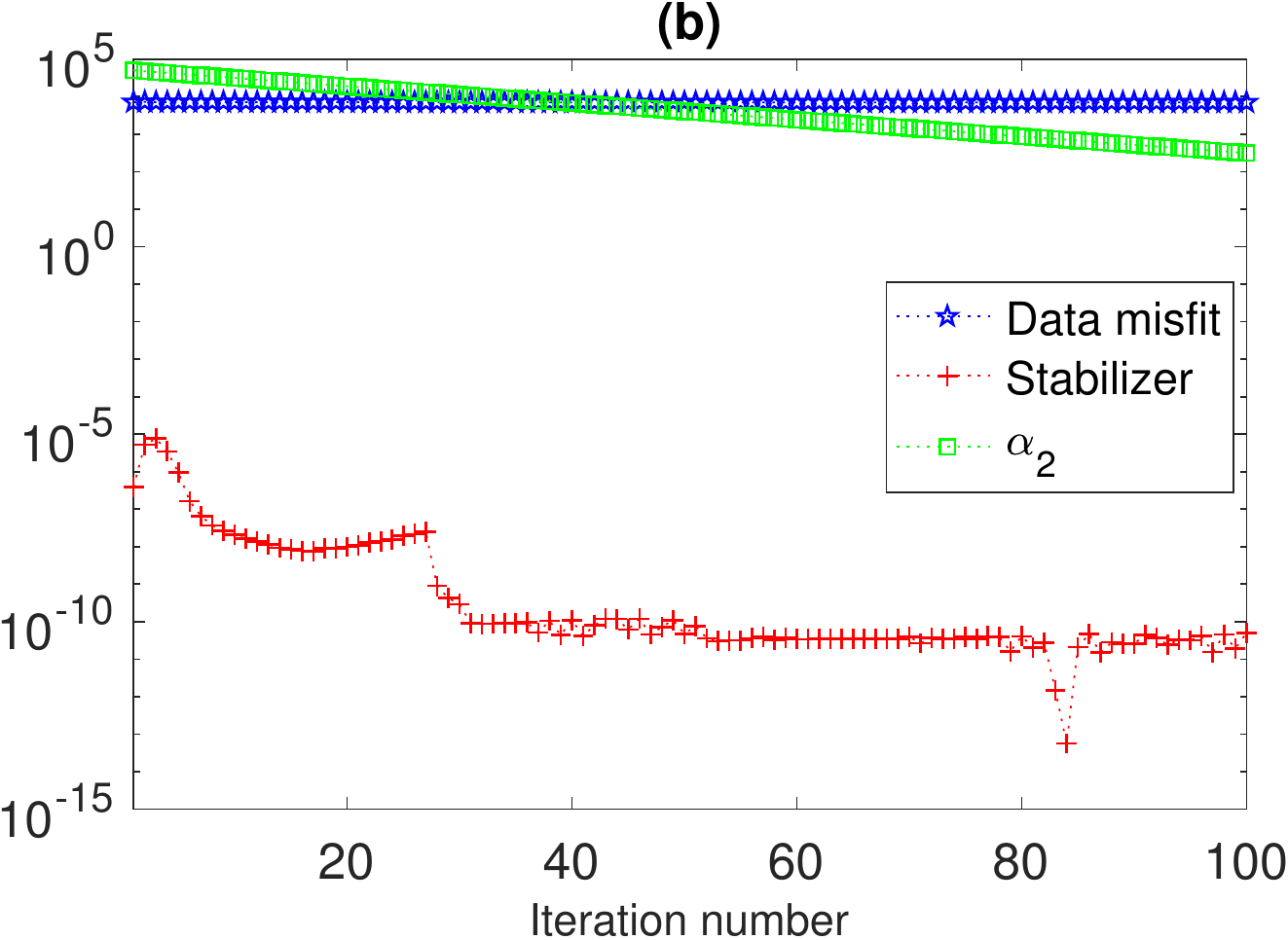}}
\caption {The progression of the data misfit, the regularization term, and the regularization parameter, with iteration  $\ell$, for the models shown in Fig.~\ref{fig9}, Model~$1$ and Case~$2$. (a) Gravity; and (b) Magnetic.}\label{fig11}
\end{figure*}

We next perform the same simulation but select the weighting on the cross-gradient to be very large, with $\lambda =10^{12}$.   In this case, the inversion terminates at  $\Kmax=100$, without satisfying the noise levels. That means that it is not possible to satisfy the data misfit criteria with a large $\lambda$. The results are presented in Figs.~\ref{fig9}, \ref{fig10}, and \ref{fig11}. The reconstructed models are not at all consistent with the original models, either with respect to the shape  or to the maximum values of the physical properties, especially for reconstruction of the density. Although, the main reconstructed bodies are quite similar to each other, as expected due to the strong requirement imposed by the use of the cross-gradient constraint, some additional unrealistic structures appear in the density model. Clearly the selection of $\lambda$ is very important. But, this is not a difficult task. It is sufficient to consider the  entries of $\bfB$, or to run the algorithm once, to determine a suitable $\lambda$.

\subsubsection{Case 3: joint L$_{1}$-norm inversion with cross-gradient constraint and hard constraint matrix}\label{Case3}
\begin{figure*}
\subfigure{\label{fig12a}\includegraphics[width=0.49\textwidth]{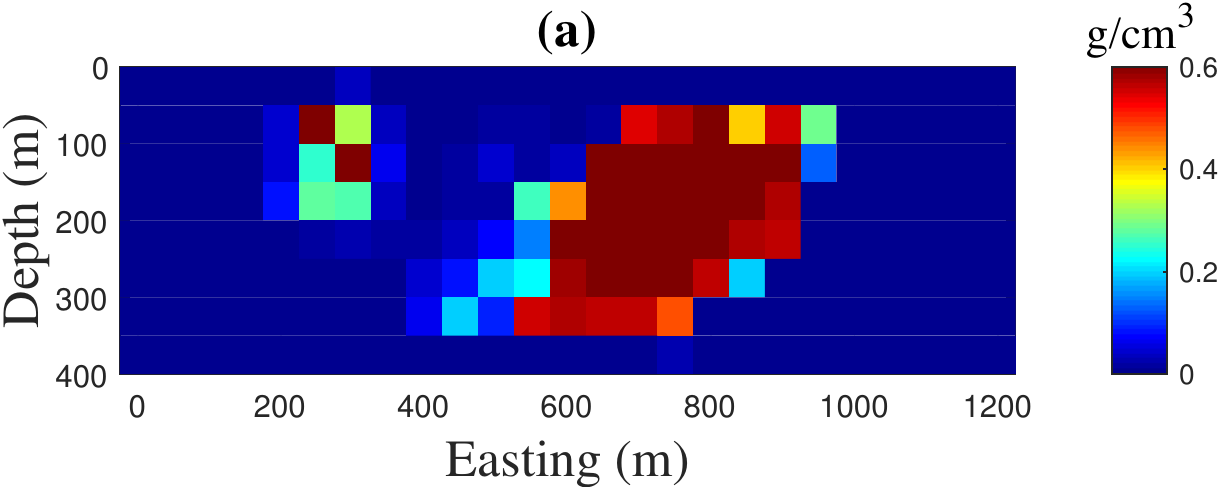}}
\subfigure{\label{fig12b}\includegraphics[width=0.49\textwidth]{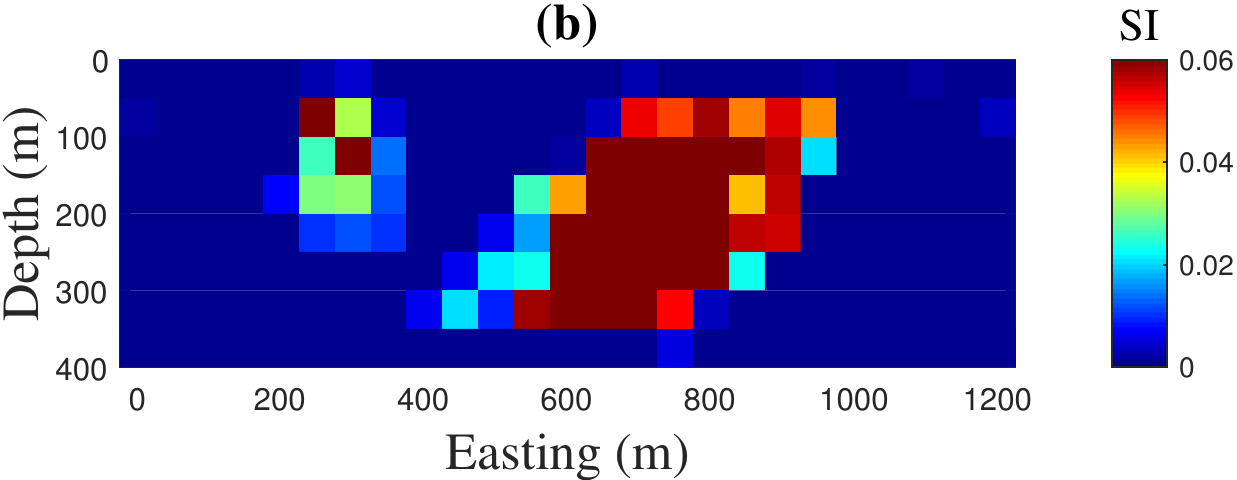}}
\caption {Reconstructed models using joint inversion with $\lambda=10^{6}$,  with the L$_{1}$-norm of the model parameters as the stabilizer,  and with the application of the  hard constraint matrix, Model~$1$ and Case~$3$. (a) Density distribution; (b) Susceptibility distribution.}\label{fig12}
\end{figure*}

The algorithm is implemented again with  $\lambda=10^{6}$, but now,  based on available information, we suppose that the physical properties for one prism in the vertical dike, located in the upper left side, are known. These known  values are communicated to the algorithm via $\bfma$ and by setting the corresponding entries of $\Wh$ to $100$. The inversion terminates at iteration $\mathrm{IT}=59<\Kmax$. The reconstructed models are presented in Figs.~\ref{fig12a} and \ref{fig12b}. The results are even better than those obtained by the previous reconstruction, as demonstrated by the model illustrated in Fig.~\ref{fig6}. Here, the known model parameters  are kept fixed during the iterations. The results demonstrate how the incorporation of available information for some model parameters can increase the reliability of the model obtained using the inverse algorithm.

\subsubsection{Case 4: joint minimum structure inversion with cross-gradient constraint}\label{Case4}
\begin{figure*}
\subfigure{\label{fig13a}\includegraphics[width=0.49\textwidth]{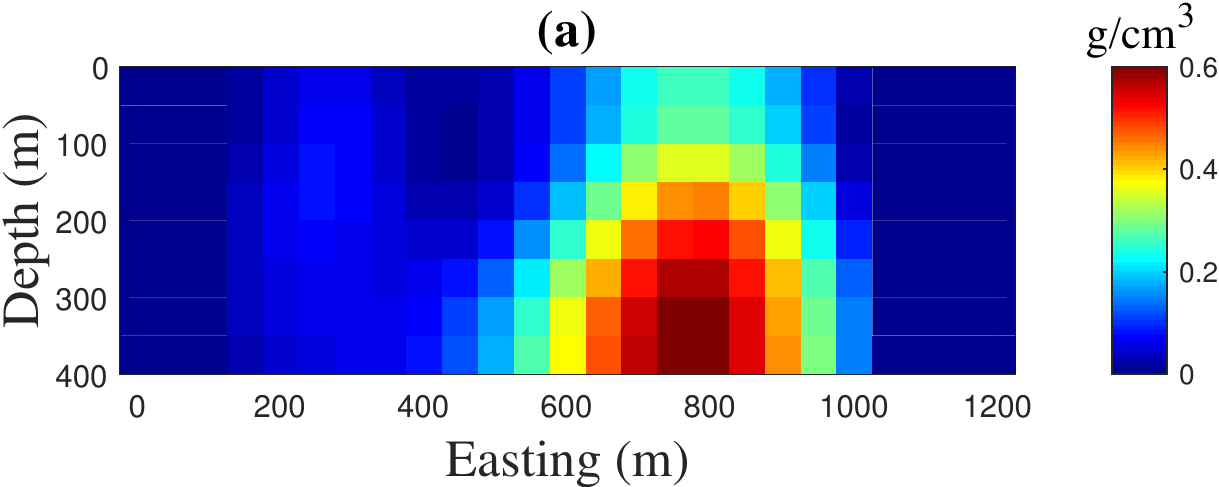}}
\subfigure{\label{fig13b}\includegraphics[width=0.49\textwidth]{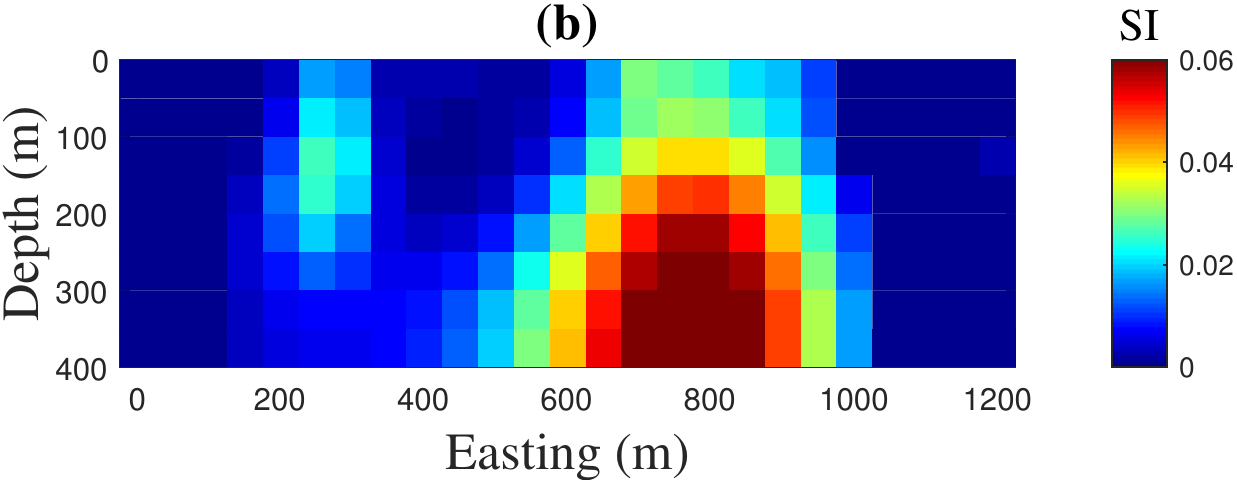}}
\caption {Reconstructed models using joint inversion with $\lambda=10^{7}$ and with the L$_{2}$-norm of the gradient of the model parameters as the stabilizer, Model~$1$ and Case~$4$. (a) Density distribution; (b) Susceptibility distribution.}\label{fig13}
\end{figure*}

\begin{figure*}
\subfigure{\label{fig14a}\includegraphics[width=0.47\textwidth]{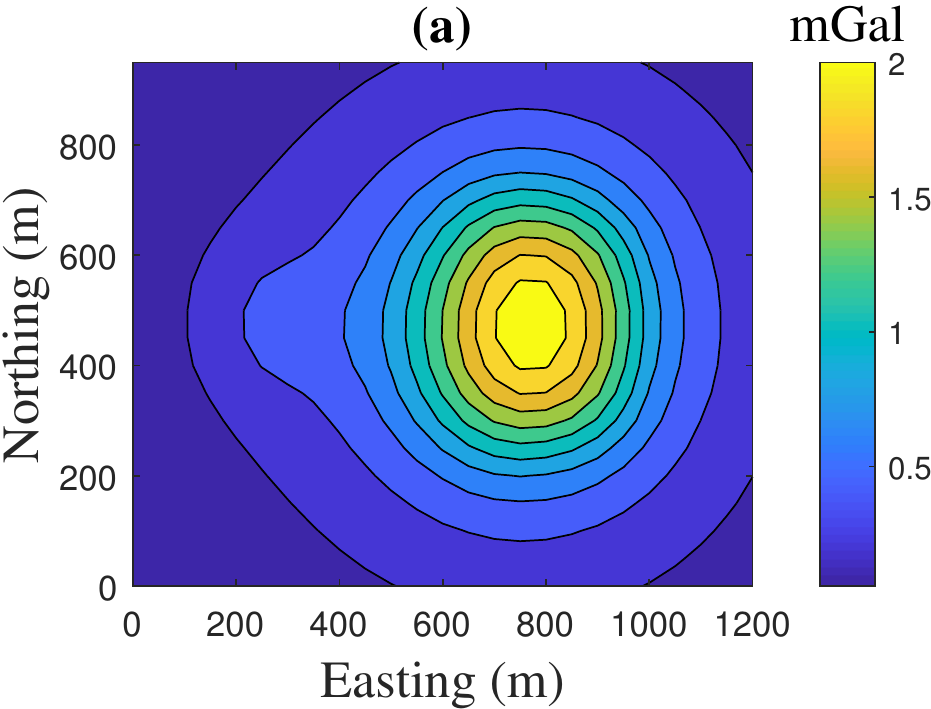}}
\subfigure{\label{fig14b}\includegraphics[width=0.48\textwidth]{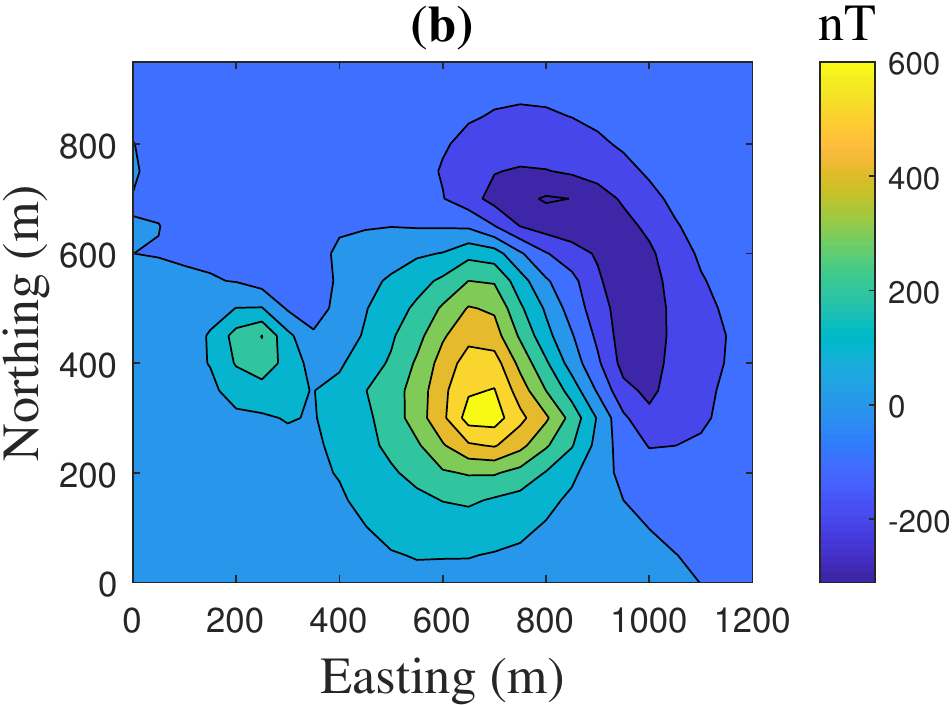}}
\caption {The data produced by the models shown in Fig.~\ref{fig13}, Model~$1$ and Case~$4$. (a) Vertical component of gravity; (b) Total magnetic field.}\label{fig14}
\end{figure*}

\begin{figure*}
\subfigure{\label{fig15a}\includegraphics[width=0.49\textwidth]{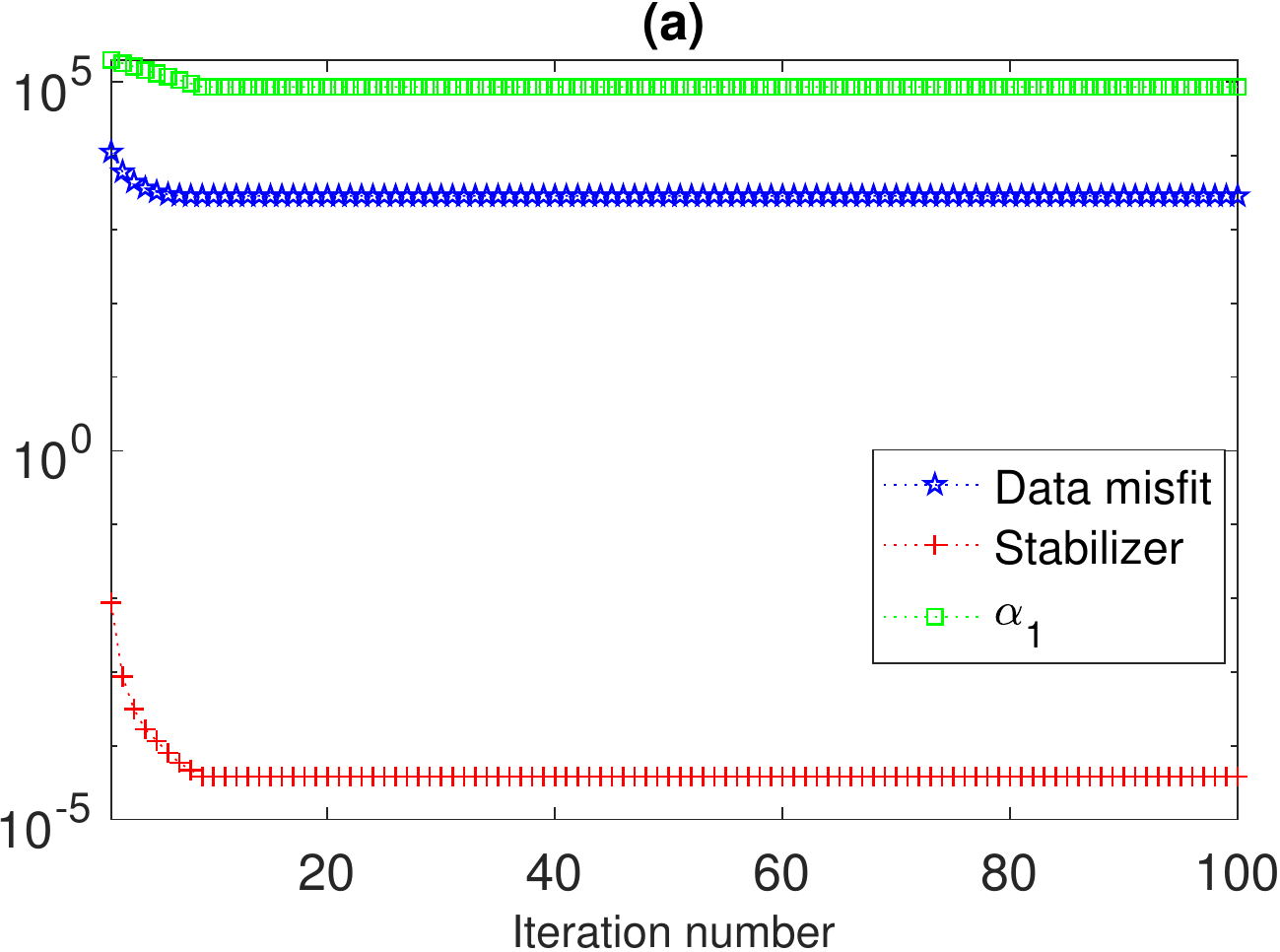}}
\subfigure{\label{fig15b}\includegraphics[width=0.49\textwidth]{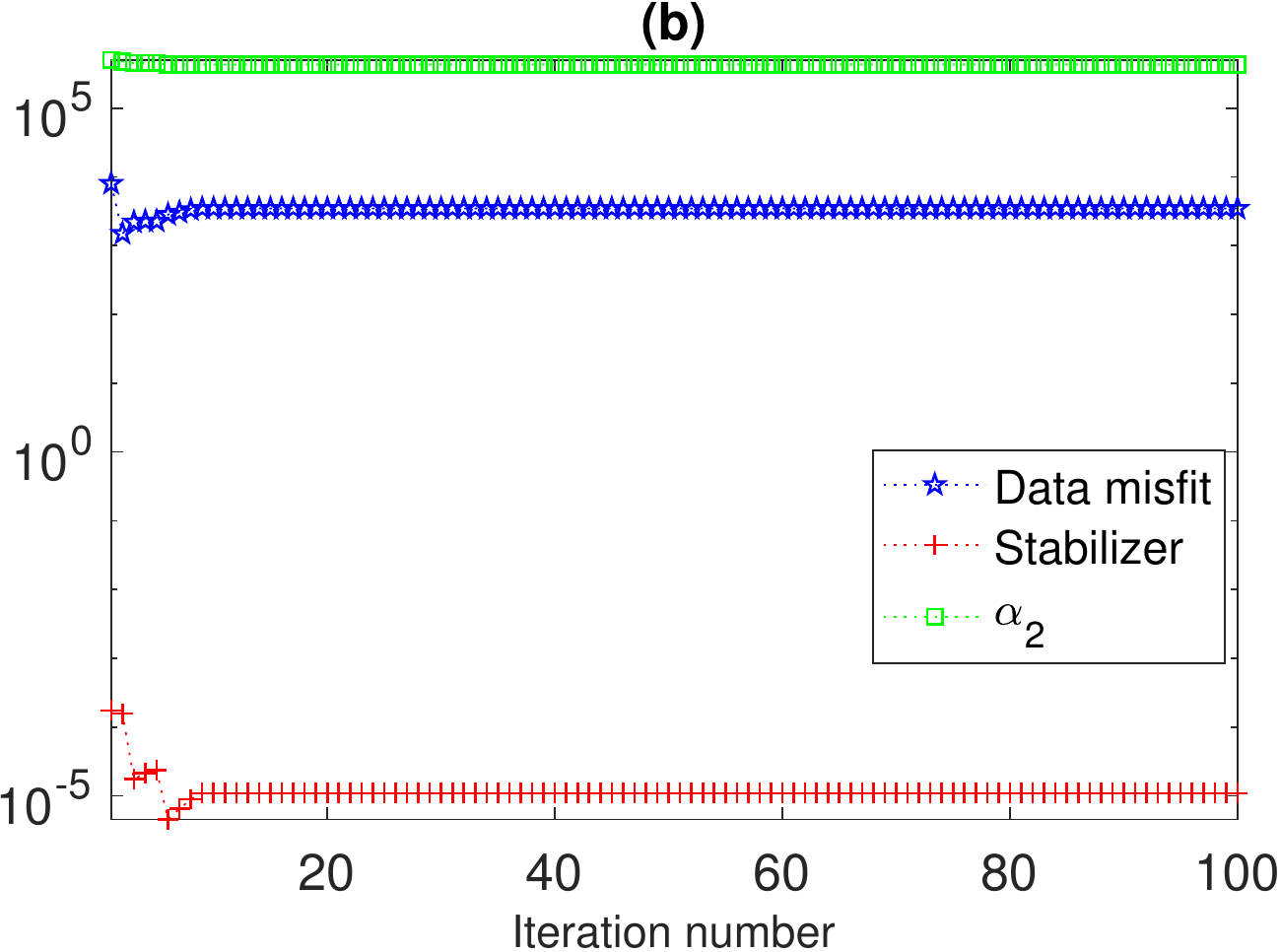}}
\caption {The progression of the data misfit, the regularization term, and the regularization parameter, with iteration  $\ell$, for the models shown in Fig.~\ref{fig13}, Model~$1$ and Case~$4$. (a) Gravity; and (b) Magnetic.}\label{fig15}
\end{figure*}

For this example, we implement the joint inversion algorithm using the minimum structure stabilizer and initial regularization parameters, $\alphag^{(1)}=200,000$ and $\alpham^{(1)}=500,000$. These values are  larger than were used for the  L$_{1}$-norm implementations discussed for Model~$1$ with Cases $1$ to $3$, in Sections~\ref{Case1}-\ref{Case3}.  This increase in the regularization parameters is  due to the significant change in the minimization function when changing the stabilizer from the  L$_{1}$-norm to  the L$_{2}$-norm.  Moreover, we also use $\lambda=10^{7}$ in order to increase the weighting on the cross-gradient constraint, consistently  with increasing the weight on the stabilizer. All other parameters remain the same as selected for Cases~$1$ to $3$.  It is also important to note that for this simulation, even though  $\alphag$ and $\alpham$ decrease as determined by $\qg$ and $\qm$, the data misfit does not necessarily decrease monotonically with the iterations.  
Because this increase in the data misfit can occur, the algorithm is modified to hold $\alphag$ or $\alpham$ fixed  for any iteration in which an increase in the calculated data misfit occurs. Moreover, the related model update for that iteration is rejected. Specifically, this means that the step is repeated without the decrease in the regularization parameter. This experience demonstrates that it is important to determine a reliable automatic parameter-choice strategy during the inversion, which is a complicated problem when three parameters are involved. The approach adopted here, which is quite simple but probably not optimal,  leads to acceptable solutions, as compared with those presented by Fregoso \& Gallardo \shortcite{FG:09}. The inversion terminates at  $\Kmax=100$. Equivalently, this means that the $\chi^2$ tests on the noise level are not satisfied for both data sets.  The results are presented in Figs.~\ref{fig13}, \ref{fig14}, and \ref{fig15}. As expected from the use of the minimum structure constraint, the reconstructed models are smooth. Moreover, as imposed by the use of the cross-gradient constraint, the models have a similar structure. They indicate a  large dipping dike  along with a small vertical structure in the subsurface. We note that the minimum structure inversion has its own advantages, which should not be disregarded, and that make the algorithm a safe strategy for the reconstruction of low-frequency subsurface structures. The progression of the data misfit, the regularization term, and the regularization parameters, with iteration  $\ell$, are presented in   Fig.~\ref{fig15}. These plots demonstrate that it is only at the initial iterations where there are significant changes in the parameters between successive iterative steps. Once the parameters are effectively fixed, the changes are very small, which suggests that the algorithm can be safely terminated for a smaller value of $\Kmax$.  For direct comparison with the presented results given for Model~$1$ with Cases $1$ to $3$ we keep $\Kmax=100$. 

\subsection{Model study 2}\label{model2}
\begin{figure*}
\subfigure{\label{fig16a}\includegraphics[width=0.49\textwidth]{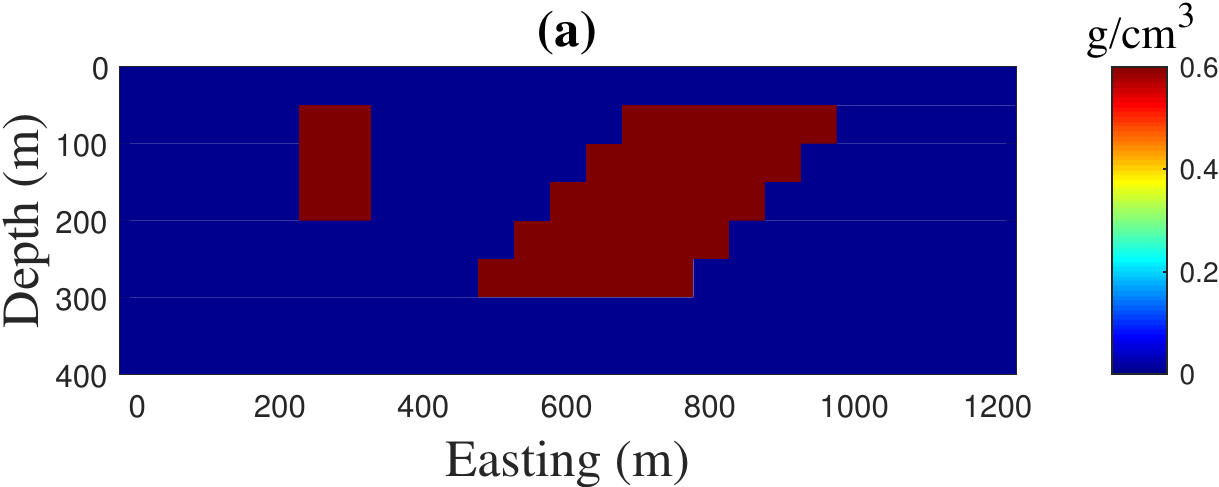}}
\subfigure{\label{fig16b}\includegraphics[width=0.49\textwidth]{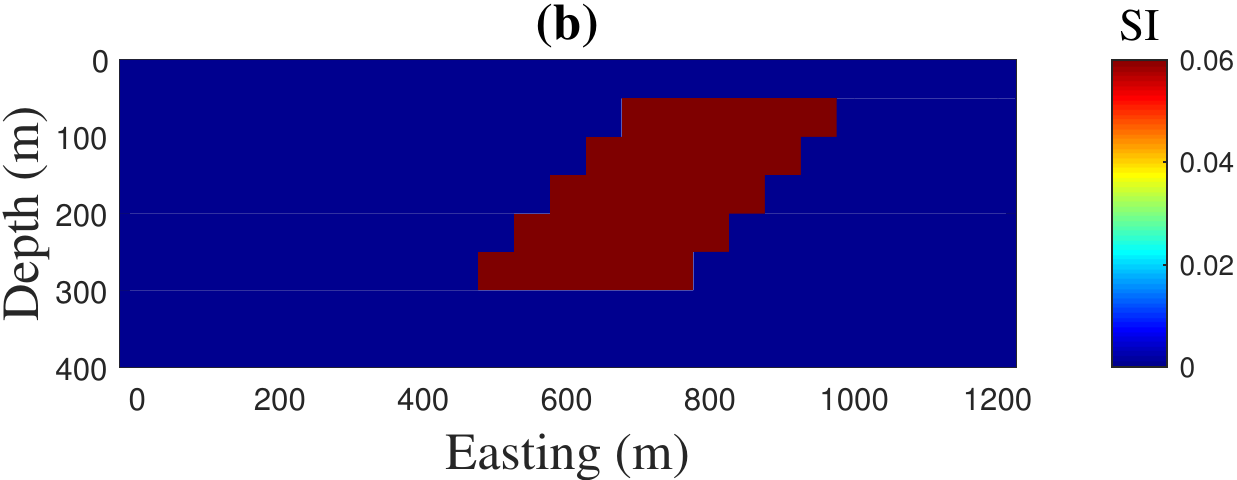}}
\caption {Cross-section of the synthetic model $2$. (a) Density distribution; (b) Susceptibility distribution. Here, it is assumed that the vertical dike is not magnetic.}\label{fig16}
\end{figure*}

\begin{figure*}
\subfigure{\label{fig17a}\includegraphics[width=0.47\textwidth]{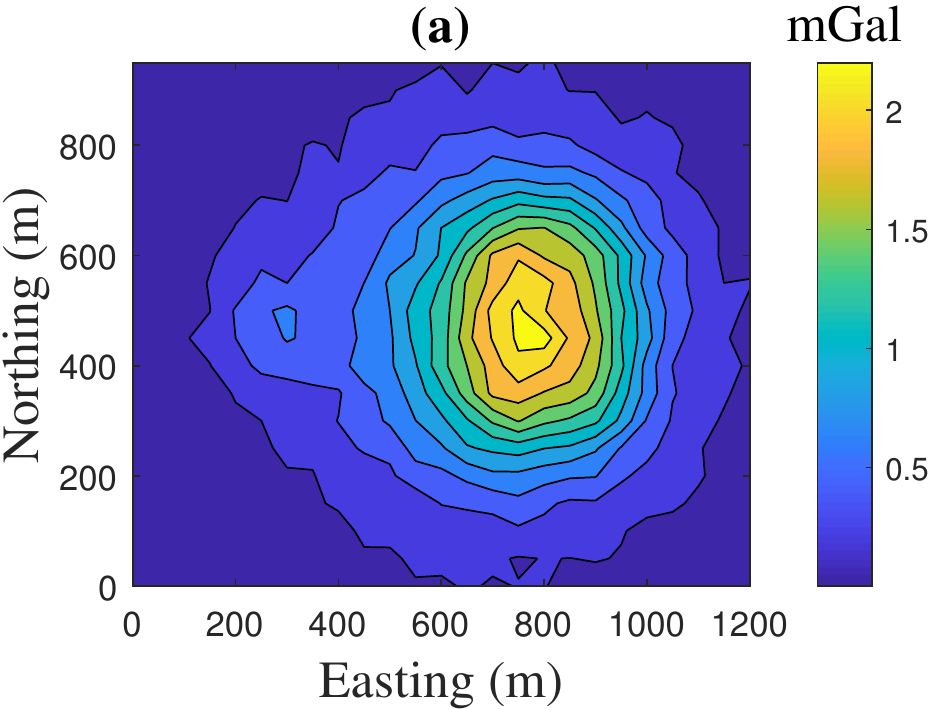}}
\subfigure{\label{fig17b}\includegraphics[width=0.48\textwidth]{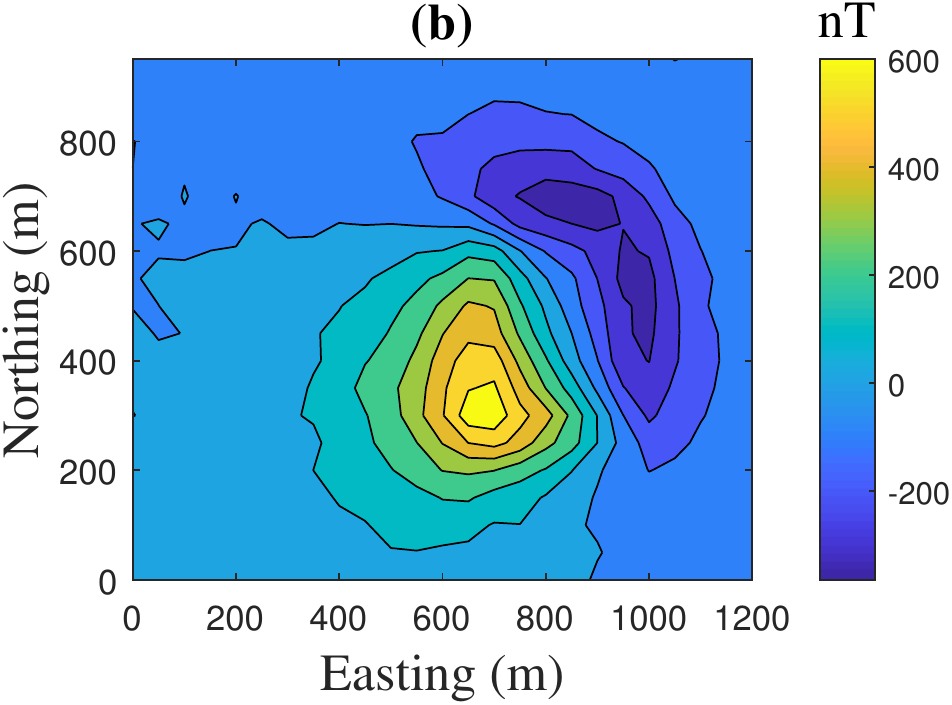}}
\caption {Noise contaminated anomaly produced by the model shown in Fig.~\ref{fig16}. (a) Vertical component of gravity; (b) Total magnetic field. The SNR for gravity and magnetic data, respectively, are $25.1778$ and $25.1773$.}\label{fig17}
\end{figure*}

For the second model we assume that the dipping dike has both a density and a susceptibility distribution, but that the vertical dike is not magnetic and has, therefore, only a  density distribution. Figs.~\ref{fig16a} and \ref{fig16b} illustrate the models. Here, the aim is to test the joint inversion algorithm, when one model has a structure in a location where the other model does not. The data produced by the models are presented in Figs.~\ref{fig17a} and \ref{fig17b}, respectively.

\subsubsection{Case 1: joint L$_{1}$-norm inversion with cross-gradient constraint}\label{Case11}
\begin{figure*}
\subfigure{\label{fig18a}\includegraphics[width=0.49\textwidth]{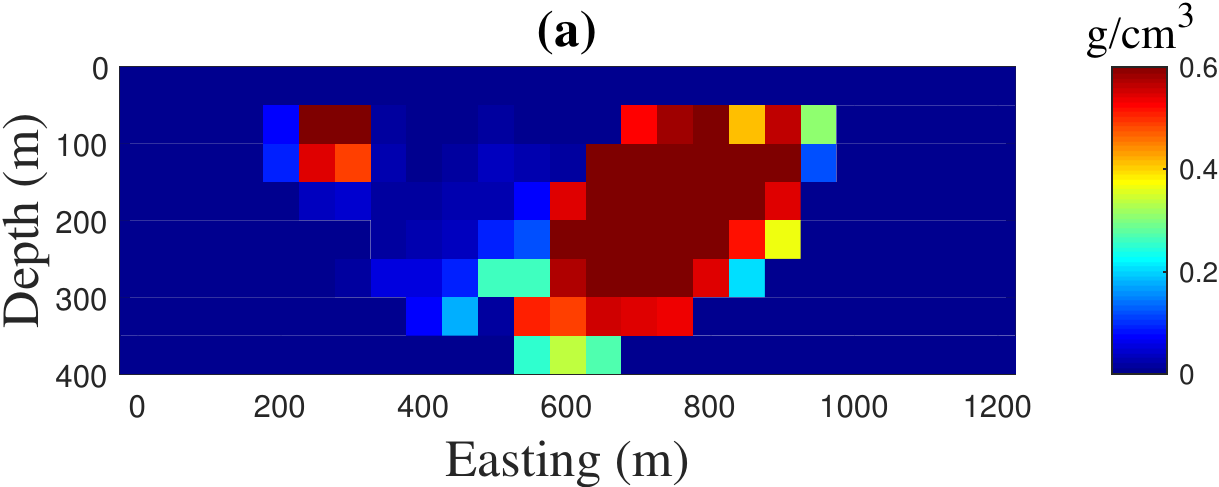}}
\subfigure{\label{fig18b}\includegraphics[width=0.49\textwidth]{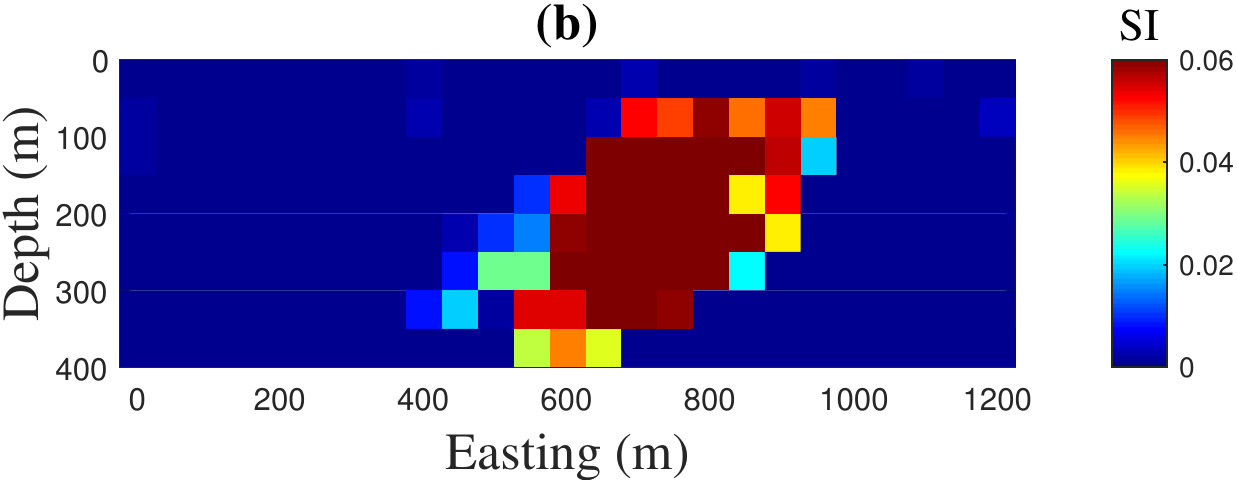}}
\caption {Reconstructed models using joint inversion with $\lambda=10^{6}$ and with the L$_{1}$-norm of the model parameters as the stabilizer, Model~$2$ and Case~$1$.  (a) Density distribution; (b) Susceptibility distribution.}\label{fig18}
\end{figure*}

The joint inversion algorithm is implemented using the L$_{1}$-norm stabilizer and with the parameters the same as were selected for the Case~$2$ simulation of Model~$1$ in Section~\ref{Case2}. The inversion terminates at  $\mathrm{IT}=58$, one iteration less than its counterpart in Section~\ref{Case2}. The reconstructed models are shown in Figs.~\ref{fig18a} and \ref{fig18b}. The dipping dikes, in both density and susceptibility distributions, are reconstructed very well, and they are completely similar. The density distribution of the vertical dike is almost reconstructed and no susceptibility distribution is obtained. This confirms, as noted by the theory of the cross-gradient constraint, that it is possible to reconstruct models which do not share all structures. Only the structures which are supported by the data will be similar.

\subsubsection{Case 2: joint minimum structure inversion with cross-gradient constraint}\label{Case22}
\begin{figure*}
\subfigure{\label{fig19a}\includegraphics[width=0.49\textwidth]{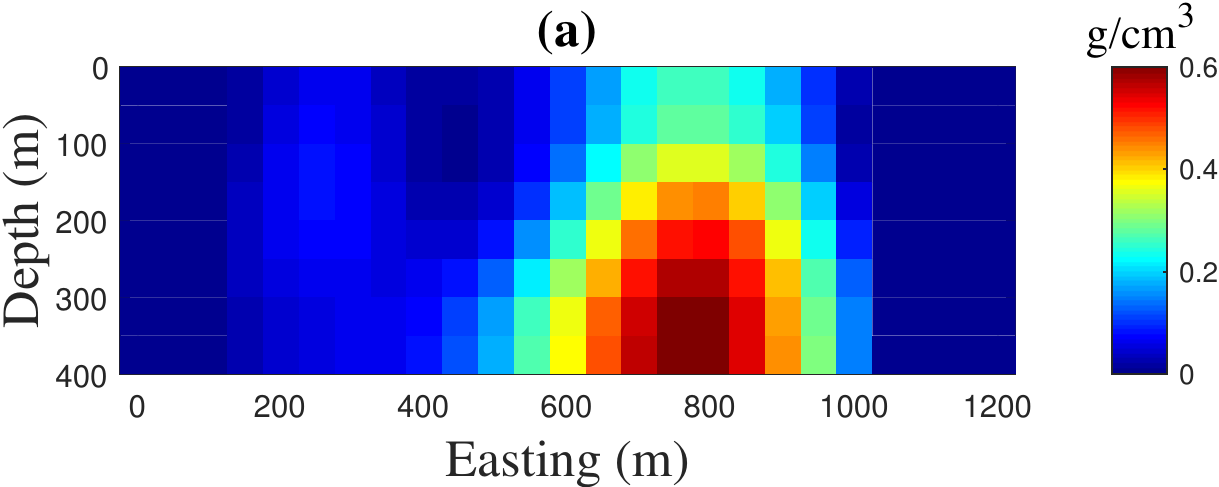}}
\subfigure{\label{fig19b}\includegraphics[width=0.49\textwidth]{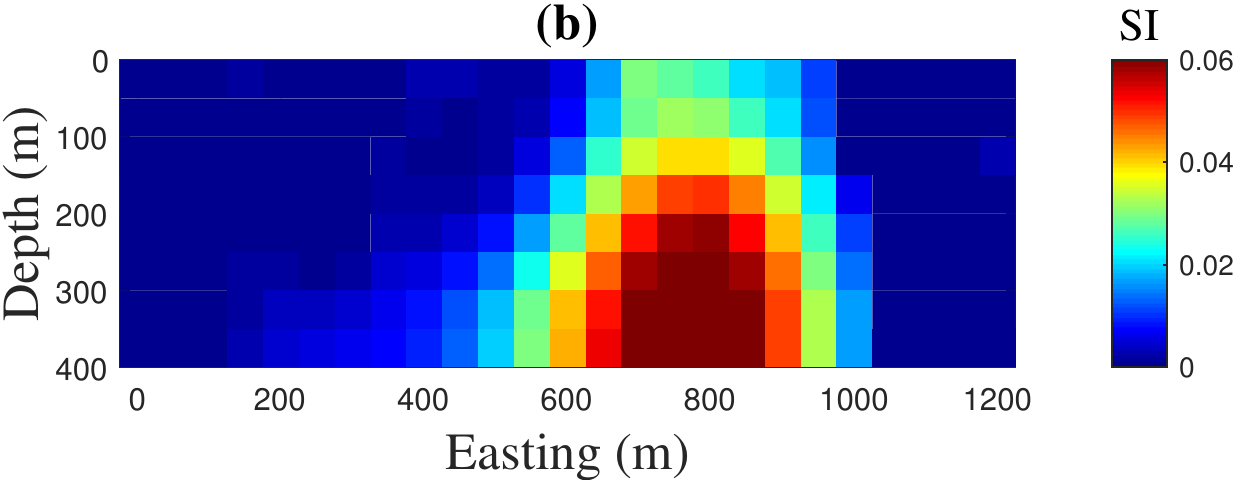}}
\caption {Reconstructed models using joint inversion with $\lambda=10^{7}$ and with the L$_{2}$-norm of the gradient of the model parameters as the stabilizer, Model~$2$ and Case~$2$.  (a) Density distribution; (b) Susceptibility distribution.}\label{fig19}
\end{figure*}

For the final validation of the algorithm, we implemented the joint inversion algorithm using the minimum structure stabilizer for the data of Fig~\ref{fig17}. All the parameters are selected as for the simulation for Model~$1$ with Case~$4$ in  Section~\ref{Case4}. The algorithm  terminated at the maximum iteration, $\mathrm{IT}=100$. The reconstructed models are presented in Figs.~\ref{fig19a} and \ref{fig19b}. As for the simulation  with the L$_{1}$-norm of the model parameters in Case~$1$, the susceptibility model exhibits none of the structure of the vertical dike.  

\section{Conclusions}\label{conclusion}
A unified framework for the incorporation of the  L$_{p}$-norm  constraint in an algorithm for joint inversion of gravity and magnetic data, in which the cross-gradient constraint provides the link between the two models, has been developed. This unifying framework shows how it is possible to  incorporate all well-known and widely used stabilizers, that are used for  potential field inversion, within a joint inversion algorithm with the cross-gradient constraint. By suitable choices of the parameter $p$ and the weighting matrix,  that define the L$_{p}$-norm constraint, it is possible to reconstruct a subsurface target exhibiting smooth, sparse, or  blocky characteristics. The  global objective function for the joint inversion consists of a data misfit term, a general form for the stabilizer, and the cross-gradient constraint.  Their contributions to the global objective function are obtained using three different regularization parameters.  A simple iterative strategy is used to convert the global non-linear objective function to a linear form at each iteration, and  the regularization weights can be adjusted at each iteration. Depth weighting and hard constraint matrices are also used in the presented inversion algorithm. These make it  possible to weight prisms at depth, and to include the known values of some prisms in the reconstructed model. Bound constraints on the model parameters may  also be imposed at each iteration. 

Results presented for two synthetic  three-dimensional models illustrate the performance of the developed algorithm. These results indicate that, when suitable regularization parameters can be estimated, the joint inversion algorithm yields suitable reconstructions of the subsurface structures. These reconstructions are improved in comparison with reconstructions obtained using independent gravity and magnetic inversions. The structures of the subsurface targets, for both density and susceptibility distributions, are quite similar and are close to the original models. A simple but practical strategy for the estimation of the regularization parameters is provided, by which large values are used at the initial step of the iteration, with a gradual decrease in subsequent iterations, dependent on selected scaling parameters for each of the imposed gravity and magnetic constraint terms.  The weight on the cross-gradient linkage constraint is chosen to balance to the three regularization terms. The results shows that this strategy is effective, particularly given the lack of any known robust methods for automatically estimating these parameters. The latter is a topic for our future study, as is the development of an improved implementation for the practical and efficient solution of three-dimensional large-scale problems and its application for real data.

\begin{acknowledgments}
Rosemary Renaut acknowledges the support of NSF grant  DMS 1913136:   ``Approximate Singular Value Expansions and Solutions of
Ill-Posed Problems".
\end{acknowledgments}


\appendix
\section{Cross-gradient formulation}\label{App1}
\begin{figure*}
\subfigure{\includegraphics[width=0.5\textwidth]{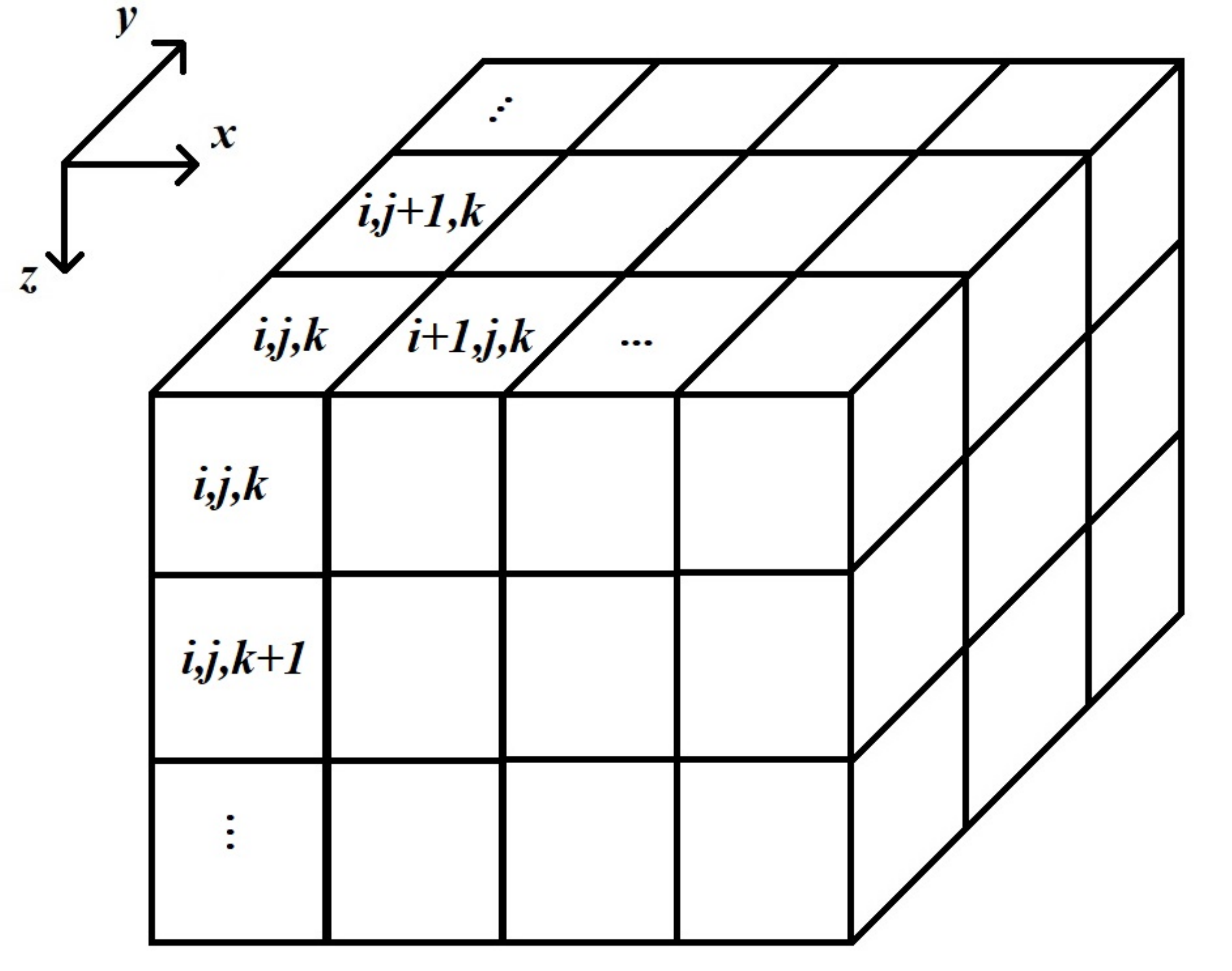}}
\caption {Discretization of the subsurface into right rectangular prisms.}\label{figA1}
\end{figure*}

The components of the cross-gradient function \eqref{crossgradientfunction} are given by 
\begin{align}\label{crosscomponent1}
 \tx(\bfmg(x,y,z),\bfmm(x,y,z))= \frac{\partial \bfmg}{\partial y} \frac{\partial \bfmm}{\partial z} - \frac{\partial \bfmg}{\partial z} \frac{\partial \bfmm}{\partial y} \in \Rm{n}\\
 \ty(\bfmg(x,y,z),\bfmm(x,y,z))= \frac{\partial \bfmg}{\partial z} \frac{\partial \bfmm}{\partial x} - \frac{\partial \bfmg}{\partial x} \frac{\partial \bfmm}{\partial z} \in \Rm{n}\\
 \tz(\bfmg(x,y,z),\bfmm(x,y,z))= \frac{\partial \bfmg}{\partial x} \frac{\partial \bfmm}{\partial y} - \frac{\partial \bfmg}{\partial y} \frac{\partial \bfmm}{\partial x} \in \Rm{n},
\end{align} 
yielding 
$
\bft(\bfm(x,y,z))= \blkstack \left( \tx,  \ty,  \tz \right) \in \Rm{3n}.
$
As illustrated in Fig.~\ref{figA1}, the subsurface is commonly divided into right rectangular prisms. Here, we suppose all prisms have same dimensions and that $\bfm(i,j,k)$ represents the value of the current estimate for $\bfm$  at  $(x, y, z)=(i\Delta x, j\Delta y, k \Delta z)$ where  $i\ge 0$, $j\ge 0$, $k\ge 0$,   with the origin, $\bfm(0,0,0)$, at the top left corner of the domain. All other parameters indexed in the same way also correspond to the parameter at the given grid point of the volume.  We use forward difference operators for each of the derivatives in \eqref{crosscomponent1} to give . 
\begin{align*}
 \tx(i,j,k)&=  \left( \frac{ \bfmg (i,j+1,k)- \bfmg(i,j,k)}{\Delta y}\right) \left(  \frac{\bfmm (i,j,k+1)- \bfmm(i,j,k)}{\Delta z}\right) -\\ &\quad \quad \left( \frac{ \bfmg (i,j,k+1)-  \bfmg(i,j,k)}{\Delta z}\right) \left(  \frac{\bfmm (i,j+1,k)- \bfmm(i,j,k)}{\Delta y}\right)\\
\ty(i,j,k)&=  \left( \frac{ \bfmg (i,j,k+1)- \bfmg(i,j,k)}{\Delta z}\right) \left(  \frac{\bfmm (i+1,j,k)- \bfmm(i,j,k)}{\Delta x}\right) -\\  &\quad \quad \left( \frac{ \bfmg (i+1,j,k)-  \bfmg(i,j,k)}{\Delta x}\right) \left(  \frac{\bfmm (i,j,k+1)- \bfmm(i,j,k)}{\Delta z}\right) \\
\tz(i,j,k)&=  \left( \frac{ \bfmg (i+1,j,k)- \bfmg(i,j,k)}{\Delta x}\right) \left(  \frac{\bfmm (i,j+1,k)- \bfmm(i,j,k)}{\Delta y}\right) -\\  &\quad \quad \left( \frac{ \bfmg (i,j+1,k)-  \bfmg(i,j,k)}{\Delta y}\right) \left(  \frac{\bfmm (i+1,j,k)- \bfmm(i,j,k)}{\Delta x}\right). 
\end{align*}
These simplify as 
\begin{align}\nonumber 
\tx(i,j,k)  =  \frac{1}{\Delta y \Delta z} \bigg(& \bfmg(i,j,k)  \Big(\bfmm (i,j+1,k) -\bfmm (i,j,k+1)\Big)  + \bfmg(i,j+1,k) \Big(\bfmm (i,j,k+1) -  \\  & \bfmm(i,j,k) \Big) +\bfmg (i,j,k+1)\Big(  \bfmm (i,j,k)- \bfmm(i,j+1,k)\Big)\bigg)\label{crosscomponent2x} \\
\ty(i,j,k)  =  \frac{1}{\Delta z \Delta x} \bigg( &\bfmg(i,j,k)  \Big(\bfmm (i,j,k+1) -\bfmm (i+1,j,k)\Big)  + \bfmg(i,j,k+1) \Big(\bfmm (i+1,j,k) -  \nonumber \\  & \bfmm(i,j,k) \Big) +\bfmg (i+1,j,k)\Big(  \bfmm (i,j,k)- \bfmm(i,j,k+1)\Big)\bigg)\label{crosscomponent2y}  \\
\tz(i,j,k)  =  \frac{1}{\Delta x \Delta y} \bigg( &\bfmg(i,j,k)   \Big(\bfmm (i+1,j,k) -\bfmm (i,j+1,k)\Big)  + \bfmg(i+1,j,k) \Big(\bfmm (i,j+1,k) -  \nonumber \\  & \bfmm(i,j,k) \Big) +\bfmg (i,j+1,k)\Big(  \bfmm (i,j,k )- \bfmm(i+1,j,k)\Big)\bigg).\label{crosscomponent2z} 
\end{align}
The Jacobian matrix for the  cross-gradient function is given by 
\begin{align}\label{JacobianApp}
\bfB=\left(\begin{array}{cc}\bfB_{\mathrm{1}x}& \bfB_{\mathrm{2} x}  \\ \bfB_{\mathrm{1}y}& \bfB_{\mathrm{2} y}   \\ \bfB_{\mathrm{1}z}& \bfB_{\mathrm{2} z}\end{array}\right) = \left(  \bfB_\mathrm{1}, \bfB_\mathrm{2} \right) \in \Rmn{3n}{2n}
\end{align}
where 
\begin{align}
\bfB_{\mathrm{i} x} = \frac{\partial \tx}{\partial \bfmi}, \quad \bfB_{\mathrm{i} y} = \frac{\partial \ty}{\partial \bfmi}, \quad \bfB_{\mathrm{i} z} = \frac{\partial \tz}{\partial \bfmi}, \mathrm{i}=1, 2.
\end{align}
We illustrate the discrete derivative for $\tx$ with respect to $\bfmg$ and $\bfmm$, first noting from \eqref{crosscomponent2x} that $\tx$ is a nonlinear combination of 
\begin{align*}
\begin{array}{ccccccc}
\bfmg(i,j,k),& \bfmg(i,j+1,k),& \bfmg(i,j,k+1) &\text{and}& \bfmm(i,j,k),& \bfmm(i,j+1,k),& \bfmm(i,j,k+1).
\end{array}
\end{align*}
Thus, component wise, there are just three derivatives with respect to each of $\bfmg$ and $\bfmm$ that are nonzero; in total there are only six non zero column entries for each row of the first row block matrix $(\bfB_{\mathrm{1} x}, \bfB_{\mathrm{2} x})$. Specifically, we only have for row $ijk$ the column entries $pqr$ as given by
\begin{align}\label{B1element}
(\bfB_{\mathrm{1} x})_{ijk, pqr}
&= \frac{1}{\Delta y \Delta z}
\left\{ \begin{array}{llll} 
\bfmm(i,j+1,k)-\bfmm(i,j,k+1)& p=i &q=j & r=k \\
\bfmm(i,j,k+1)-\bfmm(i,j,k)& p=i &q=j+1 & r=k \\
\bfmm(i,j,k)-\bfmm(i,j+1,k)& p=i &q=j & r=k+1 \\
\end{array} \right.\\
(\bfB_{\mathrm{2} x})_{ijk, pqr}
&= \frac{1}{\Delta y \Delta z}
\left\{ \begin{array}{llll} 
\bfmg(i,j,k+1)-\bfmg(i,j+1,k)& p=i &q=j & r=k \\
\bfmg(i,j,k)-\bfmg(i,j,k+1)& p=i &q=j+1 & r=k \\
\bfmg(i,j+1,k)-\bfmg(i,j,k)& p=i &q=j & r=k+1 \\
\end{array} \right.\label{B2element}
\end{align}
Here $(\bfB_{\mathrm{1}})_{ijk,pqr} $ indicates the row associated with grid point $(x_i,y_y,z_k)$, and the nonzero entries are in the relevant columns   indexed by $pqr$.
Therefore, the nonzero elements on each row  are given by   
\begin{align*}
(\bfB_{\mathrm{1}})_{ijk, \cdots} =  \frac{1}{\Delta y \Delta z}  \bigg(\cdots &\Big(\bfmm(i,j+1,k)-\bfmm(i,j,k+1)\Big) \cdots \Big(\bfmm(i,j,k+1)-\bfmm(i,j,k)\Big)  \cdots  \\
&\Big(\bfmm(i,j,k)-\bfmm(i,j+1,k)\Big) \bigg) \\
(\bfB_{\mathrm{2}})_{ijk,\cdots} =  \frac{1}{\Delta y \Delta z}  \bigg(\cdots &\Big(\bfmg(i,j,k+1)-\bfmg(i,j+1,k)\Big) \cdots \Big(\bfmg(i,j,k)-\bfmg(i,j,k+1)\Big)  \cdots  \\
&\Big(\bfmg(i,j+1,k)-\bfmg(i,j,k)\Big) \bigg).
\end{align*}
These equations are consistent with \eqref{B1element} and \eqref{B2element}. Furthermore, the non zero entries for two row block matrices associated with derivatives of $\ty$ and $\tz$ are obtained similarly from \eqref{crosscomponent2y} and \eqref{crosscomponent2z}.

\end{document}